\documentclass[11pt,a4paper]{article}
\usepackage[T1]{fontenc}
\usepackage{amsmath}
\usepackage{graphicx}
\usepackage{esint}

\makeatletter

\pdfpageheight\paperheight
\pdfpagewidth\paperwidth

\usepackage{jheppub}
\usepackage{feynman}

\title{\boldmath Holomorphic Classical Limit for Spin Effects in Gravitational and Electromagnetic Scattering }

\author[]{Alfredo Guevara}

\affiliation[]{Perimeter Institute for Theoretical Physics, Waterloo, ON N2L 2Y5, Canada}
\affiliation[]{Department of Physics $\&$ Astronomy, University of Waterloo, Waterloo, ON N2L 3G1, Canada}
\affiliation[]{CECs Valdivia $\&$ Departamento de F\'isica, Universidad de Concepci\'on, Casilla 160-C, Concepci\'on, Chile}
\emailAdd{aguevara@pitp.ca}

\abstract{We provide universal expressions for the classical piece of the amplitude given by the graviton/photon exchange between massive particles of arbitrary spin, at both tree and one loop level. In the gravitational case this leads to higher order terms in the post-Newtonian expansion, which have been previously used in the binary inspiral problem. The expressions are obtained in terms of a contour integral that computes the Leading Singularity, which was recently shown to encode the relevant information up to one loop. The classical limit is performed along a holomorphic trajectory in the space of kinematics, such that the leading order is enough to extract arbitrarily high multipole corrections. These multipole interactions are given in terms of a recently proposed representation for massive particles of any spin by Arkani-Hamed et al. This explicitly shows universality of the multipole interactions in the effective potential with respect to the spin of the scattered particles.  We perform the explicit match to standard EFT operators for $S=\frac{1}{2}$ and $S=1$. As a natural byproduct we obtain the classical pieces up to one loop for the bending of light.}

\makeatother

\begin{document}
\maketitle \flushbottom

\section{Introduction}

\label{sec:intro}

Since the early days of QFT, the use of effective methods to describe
the low energy regime of more fundamental theories \cite{HeisenbergEuler,Weinberg1,Weinberg2}
has proven extremely successful \cite{SMEFT,GiesQED-EFT,SchererChiral}.
One of the most powerful applications of Effective Field Theories
(EFTs) is the case where the high energy completion of the underlying
theory is unknown. In this direction, the problem of General Relativity
as an EFT has been studied as a tool for obtaining predictions whenever
the relevant scales are much smaller than $M_{Planck}$ \cite{Porto:16,Donoghue:17}.
For this regime the methods of QFT can be safely applied to compute
both classical and quantum long range observables. In this context,
the motivation for these problems stems from the always increasing
interest in the measurement of gravitational waves as definitive tests
of GR, which has led to the acclaimed first detection by LIGO in 2016
\cite{LIGO,LIGO2}. Specifically, the binary inspiral stage, defined
by the characteristic scale $v^{2}\sim Gm/r$, has been the subject
of extensive research since it can be addressed with analytical methods
\cite{Blanchet:02BinaryInsp,Futamase2007,Rothstein-Binary}.

\begin{figure}
\centering\includegraphics[scale=0.35]{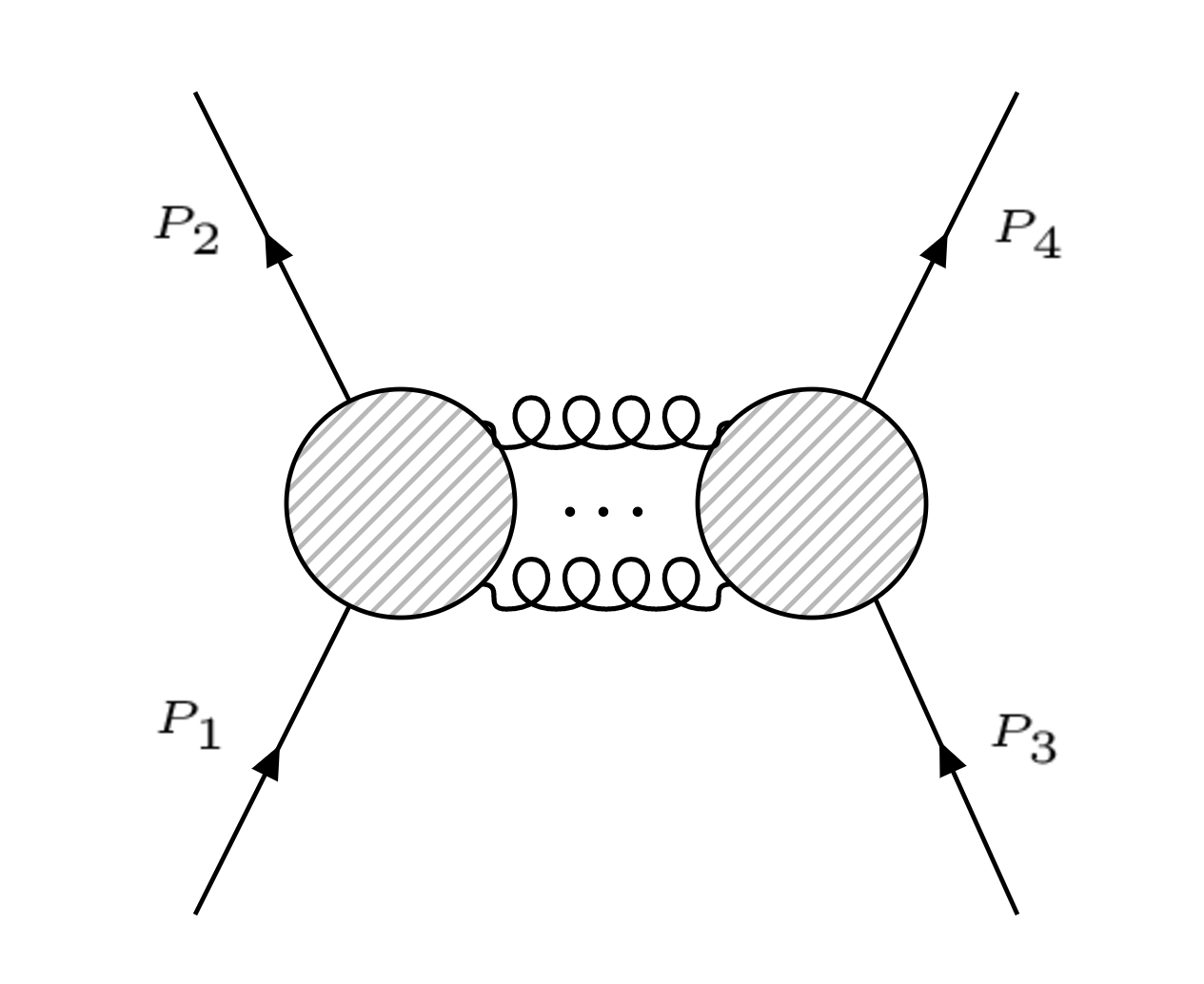}\label{fig:intro}

\caption{A typical scattering process contributing to the effective potential.
In this case two massive particles represented by $P_{1}^{2}=P_{2}^{2}=m_{a}^{2}$
and $P_{3}^{2}=P_{4}^{2}=m_{b}^{2}$ exchange several gravitons. The
momentum transfer is given by $K=P_{1}-P_{2}=(0,\vec{q})$ in the
COM frame.}
\end{figure}

The key object in the study of the binary inspiral problem is the
effective potential associated to a two-body system. This potential
admits a non-relativistic expansion in powers of $v^{2}\sim Gm/r$,
known as the post-Newtonian (PN) expansion. Pioneered by the seminal
work of Einstein-Infeld-Hoffman long ago \cite{EIH}, several attempts
have been made to evaluate the potential at higher PN orders. The
EFT approach is based on using Feynman diagrammatic techniques and
treating the PN expansion as a perturbative loop expansion \cite{Duff,Donoghue:1994dn,Muzinich:1995uj,Akhundov:1996}.
A standard setup is the $2\rightarrow2$ scattering of massive objects
$m_{a}$ and $m_{b}$, interacting through the exchange of multiple
gravitons (Fig. \ref{fig:intro}). In this case the classical potential
can be obtained from the long range behavior of the amplitude after
implementing the Born approximation \cite{Feinberg:1988yw,Holstein:2008sw,Vaidya:14}.
This classical piece is in turn extracted by setting the COM (Center
of Mass) frame, in which the momentum transfer reads $|\vec{q}|=\sqrt{-t}$
and corresponds to the Fourier conjugate of the distance $r$. Calculations
in this framework have proved extremely long and tedious, even though
there have been remarkable simplifications in the context of non-relativistic
approaches \cite{Goldberger:2004jt,Rothstein-Binary,Gilmore:2008gq,Porto:2005ac,Foffa:2016rgu}.
In addition, the electromagnetic analog of the effective potential
has been also discussed in \cite{Feinberg:1988yw,Holstein:2008sw,Holstein:2016fxh,Iwasaki}
in the context of classical corrections to Coulomb scattering. As
expected the long range behavior of this potential, i.e. the $\frac{1}{r^{n}}$
falloff, is identical to the gravitational case. The computations
are simpler in general and thus it also serves as a toy model for
the PN problem.

One of the distinctive characteristics of the PN expansion is the
treatment of the binary system as localized sources endowed with a
tower of multipole moments. The evaluation of higher multipole moments
starting at 1.5PN requires to incorporate spin into the massive particles
involved in the scattering process \cite{Porto:2005ac,Portospin2,Vaidya:14}, along with radiative corrections.
These spin contributions account for the internal angular momentum
of the objects in the macroscopic setting \cite{Barker1979,Vaidya:14}.
The universality of the gravitational coupling implies that it is
enough to consider massive particles of spin $S$ to evaluate the
spin multipole effects up to order $2S$ in the spin vector $|\vec{S}|$.
Such computation was first done up to 1-loop by Holstein and Ross
in \cite{Holstein:08} and then by Vaidya in \cite{Vaidya:14}, leading
to $|\vec{S}|^{2}$ and $|\vec{S}|^{4}$ results, respectively. The
electromagnetic counterpart has also been discussed up to $|\vec{S}|^{2}$
\cite{Holstein:2008sw}. Higher spin multipole moments are characterized
by containing higher powers of the momentum transfer $|\vec{q}|$
and $|\vec{S}|$. Thus, in order to evaluate classical spin effects
an expansion of the amplitude to arbitrarily subleading orders in
$|\vec{q}|=\sqrt{-t}$ is required. This, together with the natural
increase in difficulty for manipulating higher spin degrees of freedom
in loop QFT processes \cite{Holstein:08}, renders the computation
virtually doable only within the framework of intrinsic non-relativistic
approaches along with the aid of a computer for higher PN orders \cite{Porto:2008jj,Kol:2010ze,Levi:2014sba,Levi:2017kzq}.

In this paper we find that the combination of several new methods
can bypass some of the aforementioned difficulties. We provide fully
relativistic formulas for the classical part of the amplitude valid
for any spin at both tree and 1-loop level. The difficulty in extracting
arbitrarily subleading momentum powers is avoided by noting that the
$t\rightarrow0$ and $|\vec{q}|\rightarrow0$ expansions can be disentangled
outside the COM frame. That is, we evaluate the classical piece in
a covariant way by selecting the \textit{leading} order in the limit
$t\rightarrow0$, which we approach by using complexified momenta.
We find that the multipole terms are fully visible at leading order,
and propose Lorentz covariant expressions for them in terms of the
momentum transfer $K^{\mu}$. These expressions can then be analytically
extended to the COM frame by putting $K=(0,\vec{q})$. This is what
we call the \textit{holomorphic classical limit} (HCL).

To bypass the intrinsic complications due to the evaluation of higher
spin loop processes we draw upon a battery of modern techniques based
on the analytic structure of scattering amplitudes. In fact, techniques
such as spinor helicity formalism, on-shell recursion relations (BCFW),
and unitarity cuts have proven extremely fruitful for both computations
of gravity and gauge theory amplitudes \cite{Bern:1994cg,Bern:1994zx,Witten:2003nn,Cachazo:2004kj,Bern:2010ue,Naculich:2014naa}.
In this context, several simplifications in the computation of the
1-loop potential have already been found for scalar particles in \cite{Neill:2013wsa,Bjerrum-Bohr:2013,Holstein:2016fxh}.
Pioneered by the work of Bjerrum-Bohr et al. \cite{Bjerrum-Bohr:2014zsa}
these methods were applied to the light-bending case \cite{Bjerrum-Bohr:2014lea,Bjerrum-Bohr:2016hpa,Bai:2016ivl,Holstein:2016cfx,AmpforAstro},
where one of the external particle carries helicity $|h|\in\{0,\frac{1}{2},1\}$,
and universality with respect to $|h|$ was found. Here we extend
these approaches by considering two more techniques, both very recently
developed as a natural evolution of the previously mentioned. The
first one appeared in \cite{LS}, where Cachazo and the author proposed
to use a generalized form of unitarity cuts, known as the Leading
Singularity (LS), in order to extract the classical part of gravitational
amplitudes leading to the effective potential. It was shown that while
at tree level this simply corresponds to computing the t channel residue,
at 1-loop the LS associated to the triangle diagram leads to a fully
relativistic form containing the 1PN correction for scalar particles,
through a multidispersive treatment in the t channel. The second technique
was proposed by Arkani-Hamed et al. in \cite{Nima:16} and gives a
representation for \textit{massive} states of arbitrary spin completely
built from spinor helicity variables. Hence we use such construction
to compute the LS associated to both the gravitational and electromagnetic
triangle diagram as well as the respective tree level residues, this
time including higher spin in the external particles. The combination
of these techniques with the HCL leads to a direct evaluation of the
1-loop correction to the classical piece. The result is expressed
in a compact and covariant manner in terms of spinor helicity operators,
which are then matched to the standard spin operators of the EFT.
As a crosscheck we recover the results for both gravity and EM presented
in \cite{Holstein:08,Holstein:2008sw,Neill:2013wsa,Vaidya:14} for
$S\leq1$. By suitably defining the massless limit, we are also able
to address the light-like scattering situation and check the proposed
universality of light bending phenomena.

As an important remark, in this work we restrict our attention to
spinning particles minimally coupled to gravity or EM. This is what
is needed to reproduce the effective potential and intrinsic multipole
corrections associated to point-like sources, corresponding to black
hole processes. As a consequence we find various universalities with
respect to spin which are manifest in spinor helicity variables, and
were previously argued in \cite{Holstein:08,Bjerrum-Bohr:2013,Vaidya:14}.
The non-minimal extension, relevant for evaluating finite size effects,
is left for future work.

This paper is organized as follows. In section \ref{sec:pre} we review
the kinematics and spin considerations associated to the $2\rightarrow2$
process, which motivates the holomorphic classical limit. We then
proceed to give a short overview of the notation and conventions used
along the work, specifically those regarding manipulations of spinor
helicity variables. Next, in section \ref{sec:scalar} we review scalar
scattering and implement the HCL to extract the electromagnetic and
gravitational classical part from leading singularities at tree and
1-loop level, including the light bending case. Next, in section \ref{sec:spin}
we introduce the new spinor helicity representation for massive kinematics,
leaving the details to appendix \ref{sec:spinrep}, and use it to
extend the previous computations to spinning particles. In section
\ref{sec:discussion} we discuss the applications of these results
as well as possible future directions. Finally, in appendix \ref{sec:comframe}
we provide a prescription to match our results to the standard form
of EFT operators appearing in the effective potential for the cases
$S=\frac{1}{2},1$.

\section{Preliminaries}

\label{sec:pre}

\subsection{Kinematical Considerations and the HCL}

\label{sub:HNRL}

In the EFT framework, the off-shell effective potential can be extracted
from the S-matrix element associated to the process depicted in Fig.
\ref{fig:intro}, see e.g. \cite{Neill:2013wsa}. The standard kinematical
setup for this computation is given by the Center of Mass (COM) coordinates,
which are defined by $\vec{P_{1}}+\vec{P_{3}}=0$. We can check that
4-particle kinematics for this setup imply
\begin{equation}
(P_{1}+P_{3})\cdot(P_{1}-P_{2})=0\,,\label{eq:x-1}
\end{equation}
which means that the momentum transfer vector $K:=(P_{1}-P_{2})$
has the form 
\begin{equation}
K=(0,\vec{q})\,,\quad t=K^{2}=-\vec{q}^{2}\,,\label{eq:tcom}
\end{equation}
in the COM frame. For completeness, we also define here the average
momentum $\vec{p}$ as
\begin{equation}
\frac{P_{1}+P_{2}}{2}=(E_{a},\vec{p})\,,\quad\frac{P_{3}+P_{4}}{2}=(E_{b},-\vec{p})\,,\label{eq:average}
\end{equation}
where $E_{a},$$E_{b}$ are the respective energies for the COM frame,
while $\vec{p}^{2}\propto v^{2}$ gives the characteristic velocity
of the problem. From these definitions we can solve for the explicit
form of the momenta $P_{i}$, $i\in\{1,2,3,4\}$, and also easily
check the transverse condition $\vec{p}\cdot\vec{q}=0$. In the non-relativistic
limit $\frac{\sqrt{-t}}{m}=\frac{\text{|\ensuremath{\vec{q}}|}}{m}\rightarrow0$,
the center of mass energy $\sqrt{s}$ can be parametrized as a function
of $\vec{p}^{\,2}$. In fact, 
\begin{eqnarray}
s & = & (P_{1}+P_{3})^{2}\,,\nonumber \\
 & = & (E_{a}+E_{b})^{2}\nonumber \\
 & = & (m_{a}+m_{b})^{2}\left(1+\frac{\vec{p}^{2}}{m_{a}m_{b}}+O(\vec{p}^{\,4})\right)+O(\vec{q}^{\,2})\label{eq:nrls}
\end{eqnarray}
Note that the remaining kinematic invariant may be obtained as $u=2(m_{a}^{2}+m_{b}^{2})-t-s\,.$
In practice, we regard the amplitude for Fig. \ref{fig:intro} as
a function $M(t,s)$, which may contain poles and branch cuts in both
variables. At this point we can also introduce the spin vector $S^{\mu}$,
which will be in general constructed from polarization tensors associated
to the spinning particles, see e.g. \cite{Vaidya:14}. Suppose for
instance that the particle $m_{b}$ carries spin, then the spin vector
satisfies the transversal condition 

\begin{equation}
S^{\mu}(P_{3}+P_{4})_{\mu}=0\,,
\end{equation}
implying that in the non-relativistic regime $\vec{p}\rightarrow0$
the 4-vector becomes purely spatial, i.e. $S^{\mu}\rightarrow(0,\vec{S})$.

The PN expansion and the corresponding EM analog then proceed by extracting
the classical (i.e. $\hbar$-independent) part of the scattering amplitude
$M(t,s)$ expressed in these coordinates. This is done by selecting
the lowest order in $|\vec{q}|$ for fixed powers of $G$, spin $|\vec{S}|$
and $\vec{p}^{2}$ \cite{Neill:2013wsa,Vaidya:14}. This claim is
argued by dimensional analysis, where it is clear that for a given
order in $G$ each power of $|\vec{q}|$ carries a power of $\hbar$
unless a spin factor $|\vec{S}|$ is attached \cite{Holstein:2004dn,Porto:2005ac,Neill:2013wsa}.
Here $G$ is equivalent to 1PN order and acts as a loop counting parameter,
while the latter quantities can be counted as 1PN corrections each
\cite{Vaidya:14}. For a given number of loops and fixed value of
$s$, the expansion around $t=-\vec{q}^{2}=0$ used to select the
classical pieces coincides with the non-relativistic limit $\frac{\vec{q}}{m}\rightarrow0$.
Additionally, in the COM frame the $2^{2n}$-pole and $2^{2n-1}$-pole
interactions due to spin emerge in the form \cite{Porto:2005ac,Holstein:2008sw,Vaidya:14}
\begin{equation}
V_{S}=c_{1}(|\vec{p}|)S_{1}^{i_{1}\ldots i_{2n}}q_{i_{1}}\ldots q_{i_{2n}}+c_{2}(|\vec{p}|)S_{2}^{i_{1}\ldots i_{2n}}q_{i_{1}}\ldots q_{i_{2n-1}}p_{i_{2n}}=O(|\vec{q}|^{2n-1})\,,\label{eq:multipole}
\end{equation}
where $S_{j}^{i_{1}\ldots i_{2n}}$, $j=1,2,$ are constructed from
polarization tensors of the scattered particles in such a way that
the powers of $|\vec{S}|$ exactly match the powers of $|\vec{q}|$
in $V_{S}$. They are, in consequence, classical contributions and
correspond to the so-called mass ($j=1$) and current $(j=2$) multipoles
\cite{Vines:2016qwa}. These terms arise in the scattering amplitude
when one of the external particles, for instance the one with mass
$m_{a}$, carries spin $S_{a}\geq n$. Note that in order to evaluate
spin effects a non-relativistic expansion to arbitrary high orders
in $|\vec{q}|$ is required. To deal with this difficulty we note
that \eqref{eq:multipole} is obtained, through the non-relativistic
expansion, from the generic covariant form 

\begin{equation}
S^{\mu_{1}\cdots\mu_{m}}K_{\mu_{1}}\cdots K_{\mu_{k}}(P_{a_{k+1}})_{\mu_{k+1}}\cdots(P_{a_{m}})_{\mu_{m}}\,,\quad a_{i}\in\{1,3\}.\label{eq:covmultipole}
\end{equation}
where $k=2n$ for mass multipoles and $k=2n+1$ for current multipoles.
These spin forms are characteristic of the multipole interactions
in the sense that they are partly determined by general constraints\footnote{For instance, they vanish whenever the momentum transfer $K$ is orthogonal
to the polarization tensors $K_{\mu_{1}}\epsilon^{\mu_{1}\ldots\mu_{S}}=0$
as can be checked in \cite{Vaidya:14}, or equivalently, when it is
aligned with the spin vector.} and they emerge already in the tree level amplitude, being consistently
reproduced at the loop level \cite{Holstein:08}. We give explicit
examples of these for $S=\frac{1}{2},1$ in appendix \ref{sec:comframe}.
Once the non-relativistic limit is taken by expanding \eqref{eq:covmultipole}
with respect to $\vec{q}$ and $\vec{p}$, these terms lead to the
structures present in $V_{S}$, i.e. they capture the complete spin-dependent
couplings, together with some higher powers of $|\vec{q}|$ which
are quantum in nature. The advantage of writing the multipole terms
in the covariant form is that these are completely visible once the
limit $t=K^{2}\rightarrow0$ is taken, that is, at leading order in
the $t$ expansion. All the neglected pieces, i.e. subleading orders
in $t$, which are not captured by these multipole forms simply correspond
to quantum corrections. Thus our strategy is to compute the coefficients
associated to these EFT operators\footnote{Hereafter we may refer to the multipole terms \eqref{eq:multipole},
\eqref{eq:covmultipole} as EFT operators indistinctly. This is in
order to contrast them with the spinor operators to be defined in
section \ref{sec:spin}, which will be then matched to EFT operators.} in the $t\rightarrow0$ limit. This is done by examining the leading
order of an arbitrary linear combination of them and performing the
match with the classical piece of the amplitude, obtained by computing
the leading singularity \cite{LS}. The explicit matching procedure
is demonstrated in appendix \ref{sec:comframe}, where we use spinor
helicity variables to write the multipole terms. The idea is that
at $t=0$ the expression \eqref{eq:multipole} is not well defined
but \eqref{eq:covmultipole} is. This means that we can write our
answer for the EFT potential in terms of \eqref{eq:covmultipole}
and then proceed to analytically continue it to the region $t\neq0$,
which is easily achieved by putting $K=(0,\vec{q})$ and the corresponding
expressions for $P_{i}$. The evaluation of the classical piece near
$t=0$ is the holomorphic classical limit (HCL).

A few final remarks regarding the HCL are in order. First, as anticipated
the term holomorphic stems from the on-shell condition $P_{i}\cdot K=\pm K^{2}$,
$i\in\{1,...,4\}$, which for $t\rightarrow0$ yields $P_{i}\cdot K\rightarrow0$.
In turn, this implies that the external momenta $P_{i}$ must be complexified.
Hence, in order to reach the $t=0$ configuration we must consider
an analytic trajectory in the kinematic space, which we can parametrize
in terms of a complex variable $\beta$. We introduce such trajectory
explicitly in section \ref{sub:scalar-Loop-amplitude}, where we also
evaluate the amplitude as $\beta\rightarrow1$. Second, we stress
that just the HCL is enough to recover the classical potential with
arbitrary multipole corrections. The complete non-relativistic limit
can be further obtained by expanding around $s\rightarrow(m_{a}+m_{b})^{2}$,
i.e. expanding in $\vec{p}^{2}$ for a given power of $|\vec{q}|$.
These corrections in $\vec{p}^{\,2}$ account for higher PN corrections
when implemented through the Born approximation, which at 1-loop also
requires to subtract the iterated tree level potential. We perform
the procedure only at the level of the amplitude and refer to \cite{Feinberg:1988yw,Holstein:08,Vaidya:14,Neill:2013wsa}
for details on iterating higher PN corrections. As the expressions
we provide for the classical piece correspond to all the orders in
$\vec{p}^{\,2}$ encoded in a covariant way, we regard the HCL output
as a fully relativistic form of the classical potential. In fact,
the construction is covariant since it is based on the null condition
for $K$, which will also prove useful when defining the massless
limit of external particles for addressing light-like scattering.
Finally, the soft behavior of the momentum transfer $K\rightarrow0$,
which is the equivalent of $\frac{\vec{q}}{m}\rightarrow0$ for COM
coordinates, is not needed and we find that it does not lead to further
insights on the behavior of the potential.

\subsection{Conventions}

Before proceeding to the computation of scattering processes, we set
the conventions that will be used extensively throughout the paper.
The constructions are based on the acclaimed spinor helicity variables,
see e.g. \cite{Elvang,AmpforAstro} for a review. Here we just stress
some of the notation.

Using a mostly minus $(+,-,-,-)$ signature, a generic 4-momentum
$P^{\mu}$, with $P^{2}=m^{2}$, can be written as
\begin{equation}
P_{\alpha\dot{\alpha}}=P^{\mu}(\sigma_{\mu})_{\alpha\dot{\alpha}},\quad\bar{P}^{\dot{\alpha}\alpha}=P^{\mu}(\bar{\sigma}_{\mu})^{\dot{\alpha}\alpha}\,,
\end{equation}
where $\sigma^{\mu}=(\mathbb{I},\sigma^{i})$ and indices are lowered/raised
from the left via the $\epsilon$ tensor\footnote{Such that $\epsilon^{12}=-\epsilon_{12}=1$. },
for instance $\bar{P}^{\dot{\alpha}\alpha}=\epsilon^{\dot{\alpha}\dot{\beta}}\epsilon^{\alpha\beta}P_{\dot{\beta}\beta}$
or simply $\bar{P}=\epsilon P\epsilon^{T}$. We will also use $P$
to refer both to the 4-vector $P^{\text{\ensuremath{\mu}}}$ and the
bispinor $P_{\alpha\dot{\alpha}}$. For instance,
\begin{eqnarray}
P_{\alpha\dot{\alpha}}\bar{P}^{\dot{\alpha}\beta}=m^{2}\delta_{\alpha}^{\beta}\,, & \textsc{{\rm or}} & P\bar{P}=m^{2}\mathbb{I}\,,\label{eq:identities}\\
P_{\alpha\dot{\alpha}}\bar{Q}^{\dot{\alpha}\alpha} & ={\rm Tr}(P\bar{Q})= & 2P\cdot Q\,.\nonumber 
\end{eqnarray}
A massless momentum satisfies $\det(K)=0$ and hence can be written
as 
\begin{equation}
K_{\alpha\dot{\alpha}}=|\lambda]_{\dot{\alpha}}\langle\lambda|_{\alpha}\,,{\rm or}\,\,{\rm simply}\,\,\,K=|\lambda]\langle\lambda|.
\end{equation}
The conjugates are defined by $[\lambda|=\epsilon|\lambda]$ and $|\lambda\rangle=\langle\lambda|\epsilon^{T}$.
With these definitions $\bar{K}=|\lambda\rangle[\lambda|.$ The bilinears
$[\lambda\eta]=[\lambda|^{\dot{\alpha}}|\eta]_{\dot{\alpha}}$ and
$\langle\lambda\eta\rangle=\langle\lambda|_{\alpha}|\eta\rangle^{\alpha}$
are then naturally defined as the corresponding contractions. From
Eq. \eqref{eq:identities} we have 
\begin{equation}
[\lambda\eta]\langle\eta\lambda\rangle=2K\cdot R\,,
\end{equation}
where $R=|\eta]\langle\eta|$. This also motivates the notation

\begin{equation}
[\lambda|P|\lambda\rangle=\langle\lambda|\bar{P}|\lambda]\,,
\end{equation}
for the contraction $[\lambda|^{\dot{\alpha}}P_{\beta\dot{\alpha}}|\lambda\rangle^{\beta}$.
In the following we may omit the spinor indices $(\alpha,\dot{\alpha})$
when possible and deal with $2\times2$ operators. In appendix \ref{sec:spinrep}
we use these variables to construct the representation for massive
states of arbitrary spin, first introduced in \cite{Nima:16}.

\section{Scalar Scattering}

\label{sec:scalar}

In this section we recompute the Leading Singularity for gravitational
scattering of both tree and 1-loop level amplitudes for the no spinning
case, as first presented in \cite{LS}. This time we embed the computation
into the framework of the HCL, which will lead directly to the classical
contribution. We also present, without additional effort, the analogous
results for the EM case. Along the way we introduce new variables
which will prove helpful for the next sections. 

Let us first introduce a dimensionless variable which will be well
suited to describe the internal helicity structure of the scattering.
Motivated by the $2\rightarrow2$ process described in section \ref{sub:HNRL},
we start by considering two massive particles interacting with a massless
one. If both massive particles have the same mass $m$, the on-shell
condition for the process implies $[k|P|k\rangle=0$, where $P$ is
one of the (incoming) massive momenta and $K=|k]\langle k|$ corresponds
to the momenta of the massless particle. Thus, as proposed in \cite{Nima:16},
it is natural to introduce dimensionless variables $x$ and $\bar{x}$
such that
\begin{equation}
\begin{split}[k|P=mx\langle k|\,,\quad\langle k|\bar{P}=m\bar{x}[k|\,.\end{split}
\label{eq:3ptparam}
\end{equation}
The condition $P\bar{P}=m^{2}$ yields $x\bar{x}=1$. Note that $x$
carries helicity weight $h=+1$ under the little group transformations
of $K$. Furthermore, $mx$ precisely corresponds to the stripped
3pt amplitude for the case in which the massive particle is a scalar
and the massless particle has $h=1$\footnote{For real momenta we find that $x$ corresponds to a phase. It also
induces non-local behavior in the 3pt amplitudes \cite{Nima:16}.
However, we ignore these physical restrictions for now since we are
describing generic 3pt amplitudes which will be used to construct
the leading singularities.}. For higher helicity one simply finds ($h>0$) \cite{Nima:16}

\begin{equation}
A_{scalar}^{(+h)}=\alpha(mx)^{h}\,,\qquad A_{scalar}^{(-h)}=\alpha(m\bar{x})^{h}\,.\label{eq:3ptscalar}
\end{equation}
The (minimal) coupling constant $\alpha$ has to be chosen according
to the theory under consideration, determined once the helicity $|h|$
is given, i.e. $h=\pm1$ for EM and $h=\pm2$ for gravity. Regarding
the gravitational interaction, its universal character allows us to
fix the coupling by $\alpha=\frac{\kappa}{2}=\sqrt{8\pi G}$ irrespective
of the particle type, whereas for EM it will depend on the electric
charge carried by such particle.

\subsection{Tree Amplitude\label{sub:scalar-Tree-amplitude}}

\begin{figure}
\centering\includegraphics[scale=0.5]{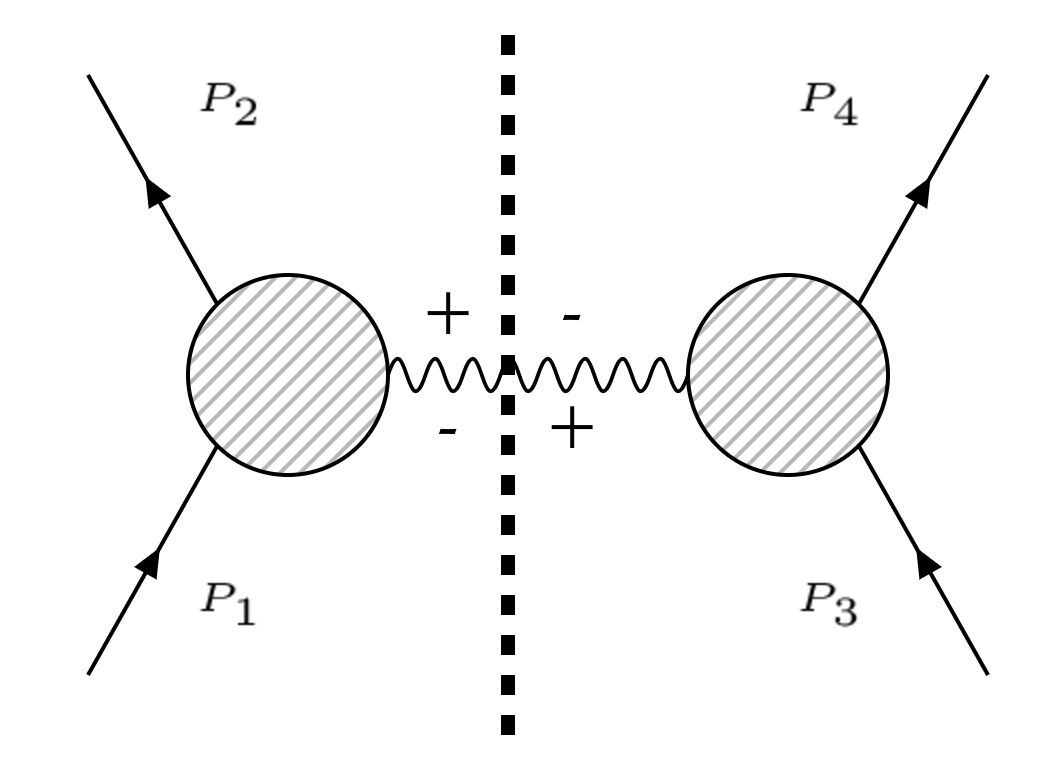}

\caption{A one photon/graviton exchange process. In the HCL the internal particle
is on-shell and the two polarizations need to be considered.\label{fig:treeex}}
\end{figure}

Let us start by computing the tree level contributions to the classical
potential. As explained in \cite{LS}, these can be directly obtained
from the Leading Singularity, which for tree amplitudes is simply
the residue at $t=0$. Here, it is transparent that the analytic expansion
around such pole will yield subleading terms $t^{n}$, $n\geq0$,
which are ultralocal (e.g. quantum) once the Fourier transform is
implemented in COM coordinates $t=-\vec{q}^{2}$ \cite{Holstein:08}.
Furthermore, by unitarity this residue precisely corresponds to the
product of on-shell 3pt amplitudes (see Fig. \ref{fig:treeex}), that
is to say, we can use the leading term in the HCL to evaluate the
classical potential. Note that, even though there exist different
couplings contributing to the s and u channel, these correspond to
contact interactions between the different particles and do not lead
to a long-range potential \cite{LS}.

With these considerations we proceed to compute the leading contribution
to the Coulomb potential by considering the one-photon exchange diagram.
Summing over both helicities and using \eqref{eq:3ptscalar} we find

\begin{equation}
M_{(0,0,1)}^{(0)}=\frac{1}{t}\left(A_{3}^{(+1)}(P_{1})A_{3}^{(-1)}(P_{3})+A_{3}^{(-1)}(P_{1})A_{3}^{(+1)}(P_{3})\right)=\alpha^{2}\dfrac{m_{a}m_{b}}{t}\left(x_{1}\bar{x}_{3}+\bar{x}_{1}x_{3}\right)\,.\label{eq:phexchange}
\end{equation}
Here we have used $M_{(S_{a},S_{b},|h|)}^{(0)}$ to denote the classical
piece of the $2\rightarrow2$ amplitude, as opposed to the notation
$A_{n}(P_{i})$ which we reserve for the $n$ pt amplitudes used as
building blocks. The index $(0)$ indicates leading order (tree level),
which will be equivalent to 0PN for the gravitational case. The subindex
$(S_{a},S_{b},|h|)=(0,0,1)$ denotes spinless particles exchanging
a photon.

The variables $x_{1}$$(\bar{x}_{1})$ and $x_{3}$$(\bar{x}_{3})$
are now associated to $P_{1}$ and $P_{3}$ respectively, through
\eqref{eq:3ptparam}. An explicit form can be obtained in terms of
the null momentum transfer $K=P_{4}-P_{3}=|k]\langle k|$, but it
is not needed here. At this stage we introduce the kinematic variables

\begin{equation}
u:=m_{a}m_{b}x_{1}\bar{x}_{3}\,,\quad v:=m_{a}m_{b}\bar{x}_{1}x_{3}.\label{eq:uvdef}
\end{equation}
Note that these variables are defined only in the HCL, i.e. for $t=0$.
Each of these carries no helicity, i.e. it is invariant under little
group transformations of the internal particle. However, they represent
the contribution from the two polarizations in the exchange of Fig.
\ref{fig:treeex}, and as such they are swapped under parity. In appendix
\ref{sec:comframe} we give explicit expressions for $u$ and $v$
in terms of their parity even and odd parts. Nevertheless, we stress
that for this and the remaining sections the only identities which
are needed can be stated as

\begin{equation}
uv=m_{a}^{2}m_{b}^{2}\,,\quad u+v=2P_{1}\cdot P_{3}\,,\label{eq:uv-1}
\end{equation}
and readily follow from their definition and \eqref{eq:3ptparam}.
We then regard the new variables as a (parity sensitive) parametrization
of the $s$ channel emerging in the HCL. Further expanding in the
non-relativistic limit yields $u,v\rightarrow m_{a}m_{b}$. 

With these definitions, we can now proceed to write the result in
a parity invariant form as

\begin{equation}
M_{(0,0,1)}^{(0)}=\alpha^{2}\dfrac{u+v}{t}=\alpha^{2}\dfrac{s-m_{a}^{2}-m_{b}^{2}}{t}\,.\label{eq:phexchange-1}
\end{equation}
After implementing COM coordinates and including the proper relativistic
normalization, this leads to the Coulomb potential in Fourier space,
which can be expanded in the limit $s\rightarrow(m_{a}+m_{b})^{2}$.
In fact, assuming both particles carry the same electric charge $e=\frac{\alpha}{\sqrt{2}}$,
we can use \eqref{eq:tcom}, \eqref{eq:nrls} to write
\begin{equation}
\frac{M_{(0,0,1)}^{(0)}}{4E_{a}E_{b}}=-\frac{e^{2}}{\vec{q}^{\,2}}\left(1+\frac{\vec{p}^{\,2}}{m_{a}m_{b}}+...\right)\,.
\end{equation}

We are now in position to easily compute the one-graviton exchange
diagram. The answer is again given by the parity invariant expression

\begin{equation}
M_{(0,0,2)}^{(0)}=\alpha^{2}\dfrac{u^{2}+v^{2}}{t}=\frac{\kappa^{2}}{4}\dfrac{(s-m_{a}^{2}-m_{b}^{2})^{2}-2m_{a}^{2}m_{b}^{2}}{t}\,.\label{eq:grexchange}
\end{equation}
Again, this leads to a relativistic expression for the Newtonian potential,
and can be put into the standard form by using the dictionary provided
in subsection \ref{sub:HNRL}
\begin{equation}
\frac{M_{(0,0,2)}^{(0)}}{4E_{a}E_{b}}=4\pi G\frac{m_{a}m_{b}}{\vec{q}^{\,2}}\left(1+\frac{(3m_{a}^{2}+8m_{a}m_{b}+3m_{b}^{2})}{2m_{a}^{2}m_{b}^{2}}\vec{p}^{\,2}+...\right),
\end{equation}
in agreement with the computations in \cite{Duff,BjerrumBohr:2002kt,Neill:2013wsa,LS}. 

Two final remarks are in order. First, it is interesting that the
gravitational result can be directly obtained by squaring the $u,v$
variables, i.e. squaring both contributions from the EM case. This
will be a general property that we will encounter again for the discussion
of the Compton amplitude in the next section, as was already pointed
out in \cite{Bjerrum-Bohr:2013} in relation with the double-copy
construction. Second, it is worth noting that up to this point no
parametrization of the external momenta was needed. In other words,
the tree level computation was done solely in terms of (pseudo)scalar
variables. As we will see now, the 1-loop case can be addressed with
the help of an external parametrization specifically designed for
the HCL. This parametrization will provide an extension of the variables
$u$ and $v$ in a sense that will become clear.

\subsection{1-Loop Amplitude: Triangle Leading Singularity\label{sub:scalar-Loop-amplitude}}

\begin{figure}
\centering\includegraphics[scale=0.35]{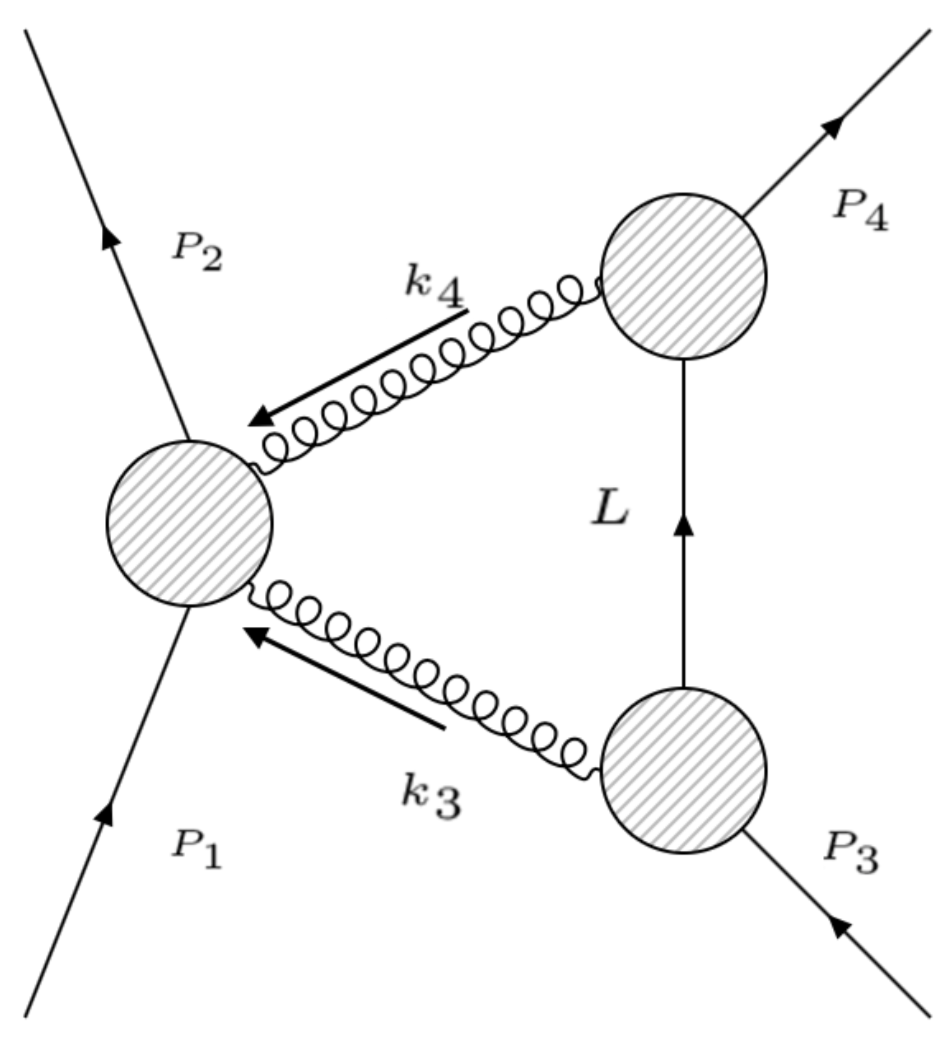}

\caption{The triangle diagram used to compute the leading singularity, corresponding
to the $b$- topology. The $a$-topology is obtained by reflection,
i.e. by appropriately exchanging the external particles. \label{fig:loopex}}
\end{figure}

Here we proceed to compute the triangle LS \cite{LS} in order to
obtain the first classical correction to the potential. This leading
singularity is associated to the 1-loop diagram arising from two photons/gravitons
exchange, Fig. \ref{fig:loopex}. As explained in the previous work,
the LS of the triangle diagram captures the second discontinuity of
the amplitude as a function of $t$, which is precisely associated
to the non-analytic behavior $\frac{1}{\sqrt{-t}}=\frac{1}{|\vec{q}|}$.
In the gravitational case this accounts for $G^{2}$ corrections or
equivalently 1PN. In order to track exclusively this contribution
we proceed to discard higher (analytic and non-analytic) powers of
$t$ by appealing to the HCL. This can be implemented to any order
in $t$ by means of the following parametrization of the external
kinematics

\begin{equation}
\begin{split}P_{3} & =|\eta]\langle\lambda|+|\lambda]\langle\eta|\,,\\
P_{4} & =\beta|\eta]\langle\lambda|+\frac{1}{\beta}|\lambda]\langle\eta|+|\lambda]\langle\lambda|\,,\\
\frac{t}{m_{b}^{2}} & =\frac{(\beta-1)^{2}}{\beta}\,,\\
\langle\lambda\eta\rangle & =[\lambda\eta]=m_{b}\,.
\end{split}
\label{eq:param}
\end{equation}
The parametrization is constructed by first defining a complex null
vector $K=|\lambda]\langle\lambda|$ orthogonal to $P_{3}$ and $P_{4}$.
Then the bispinors $(P_{3})_{\alpha\dot{\alpha}}$ and $(P_{4})_{\alpha\dot{\alpha}}$
are expanded in a suitably constructed basis, which also provides
the definition of $|\eta]_{\dot{\alpha}}$ and $\langle\eta|_{\alpha}$
up to a scale which is fixed by the fourth condition. As explained
in appendix \ref{sec:spinrep} (following the lines of \cite{Nima:16})
this basis also provides a representation for the little group associated
to massive states. The dimensionless parameter $\beta$ was called
$x$ in \cite{LS} and was introduced as a natural description of
the t channel. In this sense, this parametrization should be regarded
as an extension of the one presented there, which can be recovered
for $\beta^{2}\neq1$ by means of the shift 
\begin{equation}
|\eta]\rightarrow|\eta]+\frac{\beta}{1-\beta^{2}}|\lambda]\,,\quad\langle\eta|\rightarrow\langle\eta|-\frac{\beta}{1-\beta^{2}}\langle\lambda|\,.
\end{equation}
However, in this case we are precisely interested in the degenerate
point $\beta=1$, i.e. $t=0$, in order to define the HCL. For this
point we have $P_{4}-P_{3}=K=|\lambda]\langle\lambda|$ as the null
momentum transfer. As opposed to the tree level case, such momentum
is not associated to any particle in the exchange of Fig. \eqref{fig:loopex},
but distributed between the internal photons/gravitons. In general
for $\beta\neq1$, $K$ is just an auxiliary vector and thus we need
not to consider little group transformations for $|\lambda],\langle\lambda|$,
i.e. these are fixed spinors. Finally, we also provide a parametrization
for the $s$ channel by extending the definitions \eqref{eq:uvdef}
for $t\neq0$

\begin{equation}
\begin{split}u & =[\lambda|P_{1}|\eta\rangle\,,\quad v=[\eta|P_{1}|\lambda\rangle\,,\end{split}
\label{eq:uv}
\end{equation}
such that $u+v=2P_{1}\cdot P_{3}$ and $uv\rightarrow m_{a}^{2}m_{b}^{2}\quad\text{as}\quad\beta\rightarrow1.$ 

We are now well equipped to evaluate the triangle Leading Singularity.
Here we sketch the computation of the contour integral and refer to
\cite{LS} for further details. It is given by

\begin{equation}
\begin{split}M_{(0,0,|h|)}^{(1,b)} & =\frac{1}{4}\sum_{h_{3},h_{4}=\pm|h|}\int_{\Gamma_{{\rm LS}}}d^{4}L\,\delta(L^{2}-m_{b}^{2})\,\delta(k_{3}^{2})\,\delta(k_{4}^{2})\\
 & \times A_{4}(P_{1},-P_{2},k_{3}^{h_{3}},k_{4}^{h_{4}})\times A_{3}(P_{3},-L,-k_{3}^{-h_{3}})\times A_{3}(-P_{4},L,-k_{4}^{-h_{4}})\,,
\end{split}
\label{eq:int}
\end{equation}
where the superscript $(1,b)$ denotes the (1-loop) triangle $b$-topology
depicted in Fig. \ref{fig:loopex}. The $a$-topology is simply obtained
by exchanging particles $m_{a}$ and $m_{b}$: We leave the explicit
procedure for the appendix and in the following we deal only with
$M_{(0,0,|h|)}^{(1,b)}$. In \eqref{eq:int} we denote by $A_{3}$
and $A_{4}$ to the respective tree level amplitudes entering the
diagram (note the minus sign for \emph{outgoing }momenta), and
\begin{equation}
\begin{split}k_{3} & =-L+P_{3}\,,\quad\end{split}
k_{4}=L-P_{4}\,.\label{eq:k3k4-1}
\end{equation}
The sum is performed over propagating internal states and enforces
matching polarizations between the 3pt and 4pt amplitudes. $\Gamma_{LS}$
is a complex contour defined to enclose the emerging pole in \eqref{eq:int}.
This pole will be explicit after a parametrization for the loop momenta
$L$ is implemented and the triple-cut corresponding to the three
delta functions is performed. This will leave only a 1-dimensional
contour integral for a suitably defined $z\in\mathbb{C}$, where $L=L(z)$.
We now use the previously defined basis of spinors to parametrize
\begin{equation}
\begin{split}L(z) & =z\ell+\omega K\,,\\
\ell & =A|\eta]\langle\lambda|+B|\lambda]\langle\eta|+AB|\lambda]\langle\lambda|+|\eta]\langle\eta|\,,
\end{split}
\label{eq:L}
\end{equation}
where one scale in $\ell$ has been absorbed into $z$ and we have
further imposed the condition $\ell^{2}=0$. Using Eqs. \eqref{eq:param},
we find that implementing the triple-cut in \eqref{eq:int} fixes
$\omega(z)=-\frac{1}{z}$, while $A(z),B(z)$ become simple rational
functions of $z$ and $\beta$. The integral then takes the form 
\begin{equation}
\begin{split}M_{(0,0,|h|)}^{(1,b)}=\sum_{h_{3},h_{4}}\frac{\beta}{16(\beta^{2}-1)m_{b}^{2}} & \int_{\Gamma_{{\rm LS}}}\frac{dy}{y}A_{4}(P_{1},-P_{2},k_{3}^{h_{3}}(y),k_{4}^{h_{4}}(y))\,\\
 & \times A_{3}(P_{3},-L(y),-k_{3}^{-h_{3}}(y))\times A_{3}(-P_{4},L(y),-k_{4}^{-h_{4}}(y))\,,
\end{split}
\label{eq:int2}
\end{equation}
where $y:=-\frac{z}{(\beta-1)^{2}}$ and we now define the contour
to enclose the emergent pole at $y=\infty$, i.e. $\Gamma_{{\rm LS}}=S_{\infty}^{1}$\footnote{Also the choice $y=0$ is permitted for the contour, i.e. $\Gamma_{{\rm LS}}=S_{0}^{1}$.
This choice does not matter in the HCL since the leading piece in
\eqref{eq:int2} is invariant under the inversion of the contour \cite{LS}.} The internal massless momenta are given by

\begin{equation}
\begin{split}k_{3}(y) & =\underbrace{\frac{1}{\beta+1}\left(|\eta](\beta^{2}-1)y+|\lambda](1+\beta y)\right)}_{|k_{3}]}\,\underbrace{\frac{1}{\beta+1}\left(\langle\eta|(\beta^{2}-1)-\frac{1}{y}\langle\lambda|(1+\beta y)\right)}_{\langle k_{3}|}\,,\\
k_{4}(y) & =\underbrace{\frac{1}{\beta+1}\left(-\beta|\eta](\beta^{2}-1)y+|\lambda](1-\beta^{2}y)\right)}_{|k_{4}]}\,\underbrace{\frac{1}{\beta+1}\left(\frac{1}{\beta}\langle\eta|(\beta^{2}-1)+\frac{1}{y}\langle\lambda|(1-y)\right)}_{\langle k_{4}|}\,.
\end{split}
\label{eq:k3k4}
\end{equation}

As $\frac{\beta}{\beta^{2}-1}\rightarrow\frac{m_{b}}{2\sqrt{-t}}$
for the HCL, we find that the expression \eqref{eq:int2} already
contains the required classical correction when the leading term of
the integrand, around $\beta=1$, is extracted. We can straightforwardly
evaluate the 3pt amplitudes at $\beta=1$, giving finite contributions.
This simplification will indeed prove extremely useful for the $S>0$
cases in section \ref{sec:spin}. On the other hand, for the 4pt amplitude
the limit $\beta\rightarrow1$ is needed to obtain a finite answer,
since it contains a pole in the t channel.

Explicitly, at $\beta=1$ the internal momenta are given by 
\begin{equation}
\begin{split}k_{3}^{0}(y) & =\underbrace{\frac{1}{2}|\lambda](1+y)}_{|k_{3}]}\,\underbrace{\frac{-1}{2y}\langle\lambda|(1+y)}_{\langle k_{3}|}\,,\\
k_{4}^{0}(y) & =\underbrace{\frac{1}{2}|\lambda](1-y)}_{|k_{4}]}\,\underbrace{\frac{1}{2y}\langle\lambda|(1-y)}_{\langle k_{4}|}\,.
\end{split}
\label{eq:k3k4NRL}
\end{equation}
We thus note that in the HCL both internal momenta are collinear and
aligned with the momentum transfer $K$. For the standard non-relativistic
limit defined in the COM frame the condition $\beta\rightarrow1$
certainly implies the soft limit $K\rightarrow0$ and in general leads
to vanishing momenta for the gravitons and vanishing 3pt amplitudes
at $\beta=1$. 

Now, using the expression \eqref{eq:3ptparam} for the momenta $P_{3}$
and (outgoing) $P_{4}$, we readily find 
\begin{equation}
\begin{split}x_{3} & =x_{4}=-y\,,\end{split}
\label{eq:x3x4}
\end{equation}
such that the 3pt amplitudes are given (at $\beta=1$) by

\begin{equation}
\begin{split}A_{3}(P_{3},-L(y),-k_{3}^{+|h|}(y))A_{3}(-P_{4},L(y),-k_{4}^{-|h|}(y))\Big|_{\beta=1} & =\alpha^{2}m_{b}^{2}\,\\
A_{3}(P_{3},-L(y),-k_{3}^{+|h|}(y))A_{3}(-P_{4},L(y),-k_{4}^{+|h|}(y))\Big|_{\beta=1} & =\alpha^{2}m_{b}^{2}(y^{2})^{|h|}\,.
\end{split}
\label{eq:3ptNRL}
\end{equation}
We note that for $h_{3}=-h_{4}$ the contribution from the 3pt amplitudes
is invariant under conjugation. In fact, as can be already checked
from \eqref{eq:k3k4NRL} the conjugation is induced by $y\rightarrow-y$.
Even though the full contribution from the triangle leading singularity
requires to sum over internal helicities, in the HCL $\beta\rightarrow1$
the conjugate configuration $h_{3}=-h_{4}=-|h|$ yields the same residue,
while the configurations $h_{3}=h_{4}$ yield none as we explain below.
This means that the full result can be obtained by evaluating the
case $h_{3}=-h_{4}=+|h|$ and inserting a factor of $2$. Returning
to the computation, \eqref{eq:int2} now reads

\begin{equation}
M_{(0,0,|h|)}^{(1,b)}=\frac{\alpha^{2}}{16}\left(\frac{m_{b}}{\sqrt{-t}}\right)\int_{\infty}\frac{dy}{y}A_{(4,|h|)}^{(-+)}(\beta\rightarrow1)\,,\label{eq:int3}
\end{equation}
where $A_{(4,|h|)}^{(-+)}(\beta\rightarrow1)$ is the leading order
of the 4pt. Compton-like amplitude, given by 

\begin{equation}
A_{(4,|h|)}^{(-+)}=\alpha^{2}\begin{cases}
\dfrac{\langle k_{3}|P_{1}|k_{4}]^{2}}{\langle k_{3}|P_{1}|k_{3}]\langle k_{3}|P_{2}|k_{3}]} & |h|=1\\
\\
\frac{1}{t}\times\dfrac{\langle k_{3}|P_{1}|k_{4}]^{4}}{\langle k_{3}|P_{1}|k_{3}]\langle k_{3}|P_{2}|k_{3}]} & |h|=2
\end{cases}\label{eq:A4i}
\end{equation}
We note that the stripped Compton amplitudes \eqref{eq:A4i} exhibit
the double-copy factorization $A_{(4,2)}=4\frac{(k_{3}\cdot P_{1})(k_{3}\cdot P_{2})}{t}(A_{(4,1)})^{2}$
as explained in \cite{Bjerrum-Bohr:2013}. We will come back at this
point in section \ref{sec:spin}. By considering the definitions \eqref{eq:uv},
and using \eqref{eq:k3k4} together with momentum conservation constraints,
we find the HCL expansions
\begin{equation}
\begin{split}\langle k_{3}|P_{1}|k_{4}] & =(\beta-1)\left(u\frac{1-y}{2}+v\frac{1+y}{2}+\frac{(v-u)(1-y^{2})}{4y}\right)+O(\beta-1)^{2}\,,\\
\langle k_{3}|P_{1}|k_{3}] & =\langle k_{3}|P_{2}|k_{3}]+O(\beta-1)^{2}=(\beta-1)\frac{(v-u)(1-y^{2})}{4y}+O(\beta-1)^{2}\,.
\end{split}
\label{eq:expansion}
\end{equation}
where it is understood that $u,v$ are evaluated at $\beta=1$. We
note that the conjugation $y\rightarrow-y$ is equivalent to change
$u\leftrightarrow v$, as expected. Also, we can now argue why the
Compton amplitude gives a finite answer in the limit $\beta\rightarrow1$.
Consider for instance the gravitational case. By unitarity, such limit
induces a t channel factorization into a 3-graviton amplitude and
a scalar-scalar-graviton amplitude $A_{3}$. Because of the collinear
configuration \eqref{eq:k3k4NRL} at $\beta=1$, the 3-graviton amplitude
vanishes at the same rate as the t channel propagator $\sim(\beta-1)^{2}$,
yielding a finite result. Note that, for this factorization, regular
terms in $t$ will contribute to the result and hence these 3pt factors
are not enough to compute the HCL answer.

At this stage we exhibit for completeness the expressions for the
Compton amplitude in the case of same helicities. It is given by

\begin{equation}
A_{(++)}^{(4,|h|)}=\alpha^{2}\begin{cases}
\dfrac{[k_{3}k_{4}]^{2}}{\langle k_{3}|P_{1}|k_{3}]\langle k_{3}|P_{2}|k_{3}]} & |h|=1\\
\\
\frac{1}{t}\times\dfrac{[k_{3}k_{4}]^{4}}{\langle k_{3}|P_{1}|k_{3}]\langle k_{3}|P_{2}|k_{3}]} & |h|=2
\end{cases}\label{eq:A4i-1}
\end{equation}

By expanding $[k_{3}k_{4}]$ in an analogous form to \eqref{eq:expansion}
and, together with \eqref{eq:3ptNRL}, inserting it back into \eqref{eq:int2}
we easily find that this gives indeed vanishing residue. In fact,
this can also be checked to any order in $(\beta-1),$ i.e. with no
expansion at all \cite{LS}. As anticipated, the configurations $h_{3}=h_{4}$
simply do not lead to a classical potential.

Finally, by inserting \eqref{eq:expansion} into \eqref{eq:int3}
we find that the residue is trivial ($\text{Res}_{\infty}=1$) for
$|h|=1$, while for $|h|=2$ we have

\begin{equation}
M_{(0,0,2)}^{(1,b)}=\frac{3\alpha^{4}m_{b}}{2^{7}\sqrt{-t}}(5u^{2}+6uv+5v^{2})\,.\label{eq:int4}
\end{equation}

The expression is indeed symmetric in $u,v$, as expected by parity
invariance. By using \eqref{eq:uv-1} we can write \eqref{eq:int4-2}
in an analogous form to its tree level counterpart \eqref{eq:grexchange}

\begin{equation}
M_{(0,0,2)}^{(1,b)}=G^{2}\pi^{2}\frac{3m_{b}}{2\sqrt{-t}}\left(5(s-m_{a}^{2}-m_{b}^{2})^{2}-4m_{a}^{2}m_{b}^{2}\right)\,.\label{eq:int4-2}
\end{equation}
The contribution $M_{(0,0,2)}^{(1,a)}$ is obtained by exchanging
$m_{a}\leftrightarrow m_{b}$. After implementing the Born approximation
as explained in \cite{Feinberg:1988yw,Holstein:08}, this indeed recovers
the 1PN form of the effective potential including the corrections
in $\vec{p}^{\,2}$ \cite{BjerrumBohr:2002kt,Holstein:08,LS,Neill:2013wsa,Vaidya:14}.

\subsection{Massless Probe Particle\label{sub:Massless-probe-particle}}

Here we show that the massless case $m_{a}=0$ can be regarded as
a smooth limit defined in the variables $u,v$. In this case such
limit is natural to define since both massless and massive scalar
fields contain the same number of degrees of freedom. In appendix
\eqref{sub:Massless-representation} we show, however, how to extend
this construction to representations with nonzero spin. In the following
we focus for simplicity in the gravitational case, the electromagnetic
analog being straightforward. Moreover, the gravitational case is
motivated by the study of light bending phenomena within the framework
of EFT, see \cite{Bjerrum-Bohr:2014zsa,AmpforAstro}.

In order to discuss the massless limit, it is convenient to absorb
the mass into the definition of $x,\bar{x}$ given in \eqref{eq:3ptparam},
i.e. these quantities now carry units of energy. Then, the massless
condition $P_{3}\bar{P}_{3}=0$ is equivalent to $x_{3}\bar{x}_{3}=0$,
thus one of the helicity configurations in \eqref{eq:3ptscalar} must
vanish at $\beta=1$. This choice corresponds to selecting one of
the graviton polarizations to give vanishing contribution, that is
either $u=0$ or $v=0$. Due to parity invariance the election is
not relevant, hence we put $v=0$ and find from \eqref{eq:uv-1}

\begin{equation}
u=s-m_{b}^{2}\,,\label{eq:masslessu}
\end{equation}
which in turn yields

\begin{equation}
\begin{split}M_{m_{a}=0}^{(0)} & =\alpha^{2}\dfrac{u^{2}}{t}\\
 & =\alpha^{2}\dfrac{(s-m_{b}^{2})^{2}}{t}\,.
\end{split}
\label{eq:masslessgrex}
\end{equation}
Analogously, for the 1-loop correction \eqref{eq:int4} we find

\begin{equation}
\begin{split}M_{m_{a}=0}^{(1,b)} & =\frac{3\alpha^{4}m_{b}(5u^{2})}{2^{7}\sqrt{-t}}\,\\
 & =\frac{15\alpha^{4}}{2^{7}}\times\frac{m_{b}(s-m_{b}^{2})^{2}}{\sqrt{-t}}\,.
\end{split}
\label{eq:masslessloop}
\end{equation}
After including the normalization factor $(4E_{a}E_{b})^{-1}\approx(4E_{a}m_{b})^{-1}$
we find that this recovers the 1PN correction of the effective potential
for a massless probe particle \cite{Holstein:2016cfx,Bjerrum-Bohr:2014zsa}.
It is important to note that in this result only the $b-$topology
LS contributes, i.e. no symmetrization is needed. This is because
the triangle LS scales with the mass, i.e. for the $a-$topology is
proportional to $\frac{m_{a}}{\sqrt{-t}}$ and thus vanishes in this
case. In fact, classical effects require at least one massive propagator
entering the loop diagram \cite{Holstein:2004dn}, see also discussion.
We will again resort to this fact in section \ref{sub:lightbend},
where we construct the massless limit for spinning particles.

\section{HCL for Spinning Particles}

\label{sec:spin}

In this section we proceed to consider the case of particles with
nonzero spin. That is, we extend the computation of the triangle leading
singularity presented in section \ref{sec:scalar} but this time for
external particles with masses $m_{a}$, $m_{b}$ and spins $S_{a}$,
$S_{b}$ respectively. By using the Born approximation, the LS leads
to the 1-loop effective potential arising in gravitational or electromagnetic
scattering of spinning objects, already computed in \cite{Holstein:08}
for $S_{a},S_{b}\in\{0,\frac{1}{2},1\}$. Here we provide an explicit
expression for the tree level LS and a contour integral representation
for the 1-loop correction, both valid for any spin. We explicitly
expand the contour integral for $S_{a}\leq1$, $S_{b}$ arbitrary.
In appendix \ref{sec:comframe} we explain how to recover the results
of \cite{Holstein:08} by projecting our corresponding expression
in the HCL to the standard EFT operators.

We start by explaining a novel spinor helicity representation for
the little group of a massive particle of spin $S$, first introduced
by Arkani-Hamed et al. \cite{Nima:16}. The space is spanned by $2S+1$
polarization states, corresponding to the spin $S$ representation
of $SU(2)$. Following the lines of section \ref{sec:scalar} we will
focus on the 3pt. amplitudes $A_{3}(P_{3},P_{4},K)$ as operators
acting on in this space, which will then serve as building blocks
for the leading singularities. In our case, it will be natural to
take advantage of the parametrization of the previous section,
\begin{equation}
\begin{split}P_{3} & =|\eta]\langle\lambda|+|\lambda]\langle\eta|\,,\\
P_{4} & =\beta|\eta]\langle\lambda|+\frac{1}{\beta}|\lambda]\langle\eta|+|\lambda]\langle\lambda|\,,
\end{split}
\label{eq:param-1}
\end{equation}
to construct the little group representation for momenta $P_{3}$
and $P_{4}$ (carrying the same spin $S$) in a simultaneous fashion.
We will denote the respective $2S+1$ dimensional Hilbert spaces by
$V_{3}^{S}$ and $\bar{V}_{4}^{S}$. In appendix \ref{sec:spinrep}
we explicitly construct $V_{3}^{\frac{1}{2}}$ and $\bar{V}_{4}^{\frac{1}{2}}$
starting from the well known Dirac spinor representation. For general
spin, a basis for these spaces is given by the $2S$-th rank tensors
\footnote{The notation $|m\rangle$ for the states may seem unfortunate since
it is similar to the one for angle (chiral) spinors. However, as we
will be mostly using the anti-chiral representation for spinors, the
risk of confusion is low.}

\begin{equation}
\begin{split}|m\rangle & =\frac{1}{[\lambda\eta]^{S}}\underbrace{|\lambda]\odot\ldots\odot|\lambda]}_{m}\odot\underbrace{|\eta]\odot\ldots\odot|\eta]}_{2S-m}\,\in V_{3}^{S}\,,\\
\langle n| & =\frac{1}{[\lambda\eta]^{S}}\underbrace{[\lambda|\odot\ldots\odot[\lambda|}_{n}\odot\underbrace{[\eta|\odot\ldots\odot[\eta|}_{2S-n}\,\in\bar{V}_{4}^{S}\,,
\end{split}
\label{eq:reps}
\end{equation}
with $m,n=0,\ldots,2S$. Here the symbol $\odot$ denotes the symmetrized
tensor product. The normalization is chosen for latter convenience,
i.e.
\begin{eqnarray}
\eta_{\dot{\alpha}}\odot\lambda_{\dot{\beta}} & = & \frac{\eta_{\dot{\alpha}}\lambda_{\dot{\beta}}+\eta_{\dot{\beta}}\lambda_{\dot{\alpha}}}{\sqrt{2}}\,,\label{eq:normalization}\\
\eta_{\dot{\alpha}}\odot\lambda_{\dot{\beta}}\odot\lambda_{\dot{\gamma}} & = & \frac{\eta_{\dot{\alpha}}\lambda_{\dot{\beta}}\lambda_{\dot{\gamma}}+\eta_{\dot{\beta}}\lambda_{\dot{\alpha}}\lambda_{\dot{\gamma}}+\eta_{\dot{\gamma}}\lambda_{\dot{\alpha}}\lambda_{\dot{\beta}}}{\sqrt{3}}\,,\nonumber 
\end{eqnarray}
 etc. As we explicitly show below, in this framework we regard the
3pt amplitudes as operators $A_{S}:\,\bar{V}_{4}^{S}\otimes V_{3}^{S}\rightarrow\mathbb{\mathbb{C}}$,
that is, they are to be contracted with the states given in \eqref{eq:reps}
for both particles. The representation is symmetric and anti-chiral
in the sense that it is spanned by symmetrizations of the anti-chiral
spinors $|\lambda]$, $|\eta]$. Further details on the choice of
basis and the chirality are given in appendix \ref{sec:spinrep} (see
also \cite{Nima:16}). 

Consider then the 3pt amplitudes for two particles of momenta $P_{\text{3}}$,
$P_{4}$ and spin $S$ interacting with a massless particle of momenta
$K=P_{4}-P_{3}$ and helicity $\pm h$. From \eqref{eq:param-1} we
see that the on-shell condition $K^{2}=0$ sets $\beta=1$, i.e. $K=|\lambda]\langle\lambda|$.
For the massless particle, we choose the standard representation in
terms of the spinors $\langle k|=\frac{\langle\lambda|}{\sqrt{x}}$
and $[k|=\sqrt{x}[\lambda|$, where $x$ carries helicity weight $+1$
and agrees with the definition \eqref{eq:3ptparam} for our parametrization.
Note that $[\lambda|$ and $\langle\lambda|$ remain fixed under little
group transformations. With these conventions the minimally coupled
3pt amplitudes are given by the operators

\begin{equation}
\begin{split}A_{S}^{(+h)} & =\alpha(mx)^{h}\left(1-\frac{|\lambda][\lambda|}{m}\right)^{\otimes2S}=\alpha(mx)^{h}\left(1-\frac{|\lambda][\lambda|}{m}\right)\otimes\ldots\otimes\left(1-\frac{|\lambda][\lambda|}{m}\right)\,,\\
A_{S}^{(-h)} & =\alpha(m\bar{x})^{h}=\text{\ensuremath{\alpha}}\left(\frac{m}{x}\right)^{h}\,.
\end{split}
\label{eq:3pts}
\end{equation}
These expressions represent extensions of the ones given in \eqref{eq:3ptscalar}.
Note that we have omitted trivial tensor structures (i.e. the identity
operator) in \eqref{eq:3pts}. For example, in the second line the
explicit index structure is
\begin{equation}
\left(A_{S}^{(-h)}\right)_{\dot{\beta}_{1}\ldots\dot{\beta}_{2S}}^{\dot{\alpha}_{1}\ldots\dot{\alpha}_{2S}}=\text{\ensuremath{\alpha}}\left(\frac{m}{x}\right)^{h}\left(\mathbb{I}^{\otimes2S}\right)_{\dot{\beta}_{1}\ldots\dot{\beta}_{2S}}^{\dot{\alpha}_{1}\ldots\dot{\alpha}_{2S}}=\text{\ensuremath{\alpha}}\left(\frac{m}{x}\right)^{h}\delta_{\dot{\beta}_{1}}^{\dot{\alpha}_{1}}\ldots\delta_{\dot{\beta}_{2S}}^{\dot{\alpha}_{2S}}\,.
\end{equation}
The value for the amplitude is now obtained as the matrix element
$\langle n|A_{S}^{(\pm h)}|m\rangle$. This contraction is naturally
induced by the bilinear product $[\,,\,]$ of spinors. For instance,
consider the matrix element associated to the transition of particle
of momenta $P_{3}$ and polarization $|m\rangle$ to momenta $P_{4}$
and polarization $|n\rangle$, while absorbing a graviton: 
\begin{equation}
A^{m+(-h)\rightarrow n}=\langle n|A_{S}^{(-h)}|m\rangle=\alpha\left(\frac{m_{b}}{x}\right)^{h}\langle n|m\rangle\,,
\end{equation}
where the contraction 

\begin{equation}
\langle n|m\rangle=(-1)^{m}\delta_{m+n,2S}\,\label{eq:innerp}
\end{equation}
is induced by \eqref{eq:reps}. The relation of this contraction with
the inner product, and the corresponding normalizations, are discussed
in appendix \ref{sec:spinrep}. We note further that for helicity
$-h$ the only non trivial amplitudes are of the form $\langle n|A_{S}^{(-h)}|2S-n\rangle$
and correspond to the scalar amplitude. This is a consequence of choosing
the anti-chiral basis. For $+h$ helicity this is not the case, but
the fact that $A_{S}^{(+h)}$ is to be contracted with totally symmetric
states of $V_{3}^{S}$ and $\bar{V}_{4}^{S}$ allows us to commute
any two factors in the tensor product of \eqref{eq:3pts}. That is,
we can expand without ambiguity

\begin{eqnarray}
A_{S}^{(+h)} & = & \alpha(mx)^{h}\left(1-\frac{|\lambda][\lambda|}{m}\right)^{\otimes2S}\nonumber \\
 & = & \alpha(mx)^{h}\left(1-2S\frac{|\lambda][\lambda|}{m}+\binom{2S}{2}\frac{|\lambda][\lambda|\otimes|\lambda][\lambda|}{m^{2}}+\ldots\right)\,,\label{eq:spinexp}
\end{eqnarray}
where we again omitted the trivial operators in the tensor product.
As we explain in appendix \ref{sec:spinrep}, $|\lambda][\lambda|$
is proportional to the spin vector, hence we call it \textit{spin
operator} hereafter (see also \cite{Nima:16}). Here we can see that
in general the contraction $\langle0|A_{S}|2S\rangle$ projects out
the spin operator, again recovering the scalar amplitude.

\subsection{Tree Amplitudes}

We follow the lines of section \ref{sec:scalar} and evaluate the
$2\rightarrow2$ t channel residue. This time we assign spins $S_{a}$,
$S_{b}$ to the particles of mass $m_{a}$, $m_{b}$, respectively.
However, in order to construct the corresponding $SU(2)$ representation
\eqref{eq:reps} for the momenta $P_{1},P_{2}$, we need to repeat
the parametrization for $P_{3}$ and $P_{4}$ given in \eqref{eq:param-1}.
This time we have

\begin{equation}
\begin{split}P_{1} & =|\hat{\eta}]\langle\hat{\lambda}|+|\hat{\lambda}]\langle\hat{\eta}|\,,\\
P_{2} & =\beta|\hat{\eta}]\langle\hat{\lambda}|+\frac{1}{\beta}|\hat{\lambda}]\langle\hat{\eta}|+|\hat{\lambda}]\langle\hat{\lambda}|\,,
\end{split}
\label{eq:param2}
\end{equation}
together with the normalization $[\hat{\lambda}\hat{\eta}]=m_{a}$.
Both parametrizations can be matched in the HCL, effectively reducing
the apparent degrees of freedom. In fact, $\beta\rightarrow1$ yields
$|\lambda]\langle\lambda|\rightarrow-|\hat{\lambda}]\langle\hat{\lambda}|$.
Recall that at $\beta=1$ the tree level process of Fig. $\ref{fig:treeex}$
consists of a photon/graviton exchange, with corresponding momentum
$K=|\lambda]\langle\lambda|$. For this internal particle we choose
the spinors

\begin{equation}
|K]=|\hat{\lambda}]=\frac{|\lambda]}{\gamma}\,,\,|K\rangle=|\hat{\lambda}\rangle=-\gamma|\lambda\rangle\,,\label{eq:compat}
\end{equation}
for some $\gamma\in\mathbb{C}$. By using the definitions $\eqref{eq:3ptparam}$
for both $P_{1}$ and $P_{3}$ we find $x_{1}=1\,,\quad\bar{x}_{3}=-\gamma^{2}\,$,
Using $\eqref{eq:uvdef}$ we can then solve for $\gamma$, completely
determining $|\hat{\lambda}]$ and $\langle\hat{\lambda}|$:
\begin{equation}
\gamma^{2}=-\frac{u}{m_{a}m_{b}}=-\frac{m_{a}m_{b}}{v}.
\end{equation}

After this detour, we are ready to compute the tree level residue.
The $2\rightarrow2$ amplitude is here regarded as the operator $M_{(S_{a},S_{b},|h|)}^{(0)}:V_{1}^{S_{a}}\otimes\bar{V}_{2}^{S_{a}}\otimes V_{3}^{S_{b}}\otimes\bar{V}_{4}^{S_{b}}\rightarrow\mathbb{C}$,
where $V_{1}^{S_{a}},\bar{V}_{2}^{S_{a}}$ are constructed in analogous
manner to \eqref{eq:reps}. Using the expansion \eqref{eq:spinexp}
we find our first main result
\begin{equation}
\begin{split}M_{(S_{a},S_{b},|h|)}^{(0)} & =\alpha^{2}\dfrac{(m_{a}m_{b})^{h}}{t}\left((x_{1}\bar{x}_{3})^{h}\left(1-\frac{|\hat{\lambda}][\hat{\lambda}|}{m_{a}}\right)^{2S_{a}}+(\bar{x}_{1}x_{3})^{h}\left(1-\frac{|\lambda][\lambda|}{m_{b}}\right)^{2S_{b}}\right)\,\\
 & =\dfrac{\alpha^{2}}{t}\left(u^{h}\left(1-\frac{|\hat{\lambda}][\hat{\lambda}|}{m_{a}}\right)^{2S_{a}}+v^{h}\left(1-\frac{|\lambda][\lambda|}{m_{b}}\right)^{2S_{b}}\right)\\
 & =\dfrac{\alpha^{2}}{t}\left(u^{h}-2u^{h}S_{a}\frac{|\hat{\lambda}][\hat{\lambda}|}{m_{a}}\otimes\mathbb{I}_{b}+S_{a}(2S_{a}-1)\frac{|\tilde{\lambda}]|\hat{\lambda}][\hat{\lambda}|[\tilde{\lambda|}}{m_{a}^{2}}\otimes\mathbb{I}_{b}\right.\\
 & \quad\left.+v^{h}-2v^{h}S_{b}\,\mathbb{I}_{a}\otimes\frac{|\lambda][\lambda|}{m_{b}}+\ldots\right)\,,
\end{split}
\label{eq:spinexchange}
\end{equation}
where $h=1$ for Electromagnetism and $h=2$ for Gravity. In the third
and fourth line we exhibited explicitly the identity operators for
both representations to emphasize that the spin operators act on different
spaces and hence cannot be summed. In appendix \ref{sec:spinrep}
it is argued, by examining the $S=\frac{1}{2}$ and $S=1$ case, that
the binomial expansion is in direct correspondence with the expansion
in multipoles moments and hence to the PN expansion for the gravitational
case. That is to say we can match the operators $|\hat{\lambda}][\hat{\lambda}|^{\otimes2n}$,
$|\hat{\lambda}][\hat{\lambda}|^{\otimes2n-1}$ to the spin operators
\eqref{eq:covmultipole} in the HCL and compute the respective coefficients
in the EFT expression, as we demonstrate in appendix \ref{sec:comframe}
for the cases in the literature, i.e. $S\leq1$. Note further that
we can easily identify universal multipole interactions as predicted
by \cite{Holstein:08,Bjerrum-Bohr:2013} for the minimal coupling,
the leading one corresponding to scalar (orbital) interaction. Here
we emphasize again that all these multipole interactions can be easily
seen at $\beta=1$, in contrast with the COM frame limit.

Finally, note that the parametrization that we introduced did not
seem relevant in order to obtain \eqref{eq:spinexchange}. However,
it is indeed implicit in the choice of basis of states needed to project
the operator $M_{(S_{a},S_{b},h)}^{(0)}$ into a particular matrix
element. Next we compute the 1-loop correction for this process, which
requires extensive use of the parametrization.

\subsection{1-Loop Amplitude}

\label{sub:spin1loop}

We now compute the triangle LS \eqref{eq:int} for the case in which
the external particles carry spin. We explicitly expand the contour
integral in the HCL for the case $S_{a}\leq1$ and $S_{b}$ arbitrary.
The limitation for $S_{a}$ simply comes from the fact that for $S_{a}\leq1$
the four point Compton amplitude has a well known compact form \cite{Bjerrum-Bohr:2013}
both for EM and gravity. Let us remark that the expression for higher
spins is also known in terms of the new spinor helicity formulation
\cite{Nima:16}, but we will leave the explicit treatment for future
work. Additionally, the case $S_{a}\leq1$ is enough to recover all
the 1-loop results for the scattering amplitude in the literature
\cite{Holstein:2008sw,Holstein:08}, and suffices here to demonstrate
the effectiveness of the method (see appendix \ref{sec:comframe}).
Note that the final result is obtained by considering the two triangle
topologies for the leading singularity, which can be obtained by symmetrization
as we explain below.

In the following we regard the 3pt and 4pt amplitudes entering the
integrand \eqref{eq:int} as $2\times2$ operators equipped with the
natural multiplication. Analogous to the scalar case, only the opposite
helicities contribute to the residue and both configurations give
the same contribution, hence we focus only on $(+-)$. Furthermore,
the 3pt amplitudes can also be readily obtained at $\beta=1$, by
using \eqref{eq:k3k4NRL} into \eqref{eq:3pts}. They give

\begin{equation}
\begin{split}A_{3}(P_{3},-L(y),k_{3}^{+i}(y))A_{3}(-P_{4},L(y),k_{4}^{-i}(y))\Big|_{\beta=1} & =\alpha^{2}m_{b}^{2}\left(1-\frac{|k_{3}][k_{3}|}{ym_{b}}\right)^{2S_{b}}\,,\\
 & =\alpha^{2}m_{b}^{2}\left(1-\frac{(1+y)^{2}}{4y}\frac{|\lambda][\lambda|}{m_{b}}\right)^{2S_{b}}\,.
\end{split}
\label{eq:3ptNRLspin}
\end{equation}
This time note that the $y$ variable carries helicity weight $+1$,
as can be seen from plugging $k_{3}$ and $P_{3}$ in \eqref{eq:3ptparam}.
This means that we needed to restore the helicity factor $y$ in the
first line in order to account for little group transformations of
$k_{3}$. As in the tree level case, eq. \eqref{eq:3ptNRLspin} corresponds
to an expansion in terms of spin structures that \textquotedbl{}survive\textquotedbl{}
the limit $\beta=1$.

We now proceed to compute the 4pt Compton amplitude in the limit $\beta\rightarrow1$.
For this, consider 

\begin{equation}
A_{(4,|h|)}^{(S_{a})}=\alpha^{2}\begin{cases}
\Gamma^{\otimes2S_{a}}\dfrac{\langle k_{3}|P_{1}|k_{4}]^{2-2S_{a}}}{\langle k_{3}|P_{1}|k_{3}]\langle k_{3}|P_{2}|k_{3}]} & |h|=1,\,\\
\\
\Gamma^{\otimes2S_{a}}\frac{1}{t}\times\dfrac{\langle k_{3}|P_{1}|k_{4}]^{4-2S_{a}}}{\langle k_{3}|P_{1}|k_{3}]\langle k_{3}|P_{2}|k_{3}]} & |h|=2,
\end{cases}\label{eq:A4spin}
\end{equation}
for $S_{a}\in\{0,\frac{1}{2},1\}$. Here we have defined the $2\times2$
matrix \cite{Nima:16}

\begin{equation}
\Gamma=|k_{4}]\langle k_{3}|P_{1}+P_{2}|k_{3}\rangle[k_{4}|\,.\label{eq:Gamma}
\end{equation}
As anticipated, the 4pt. amplitude takes a compact form for $S_{a}\leq1$,
and exhibits remarkable factorizations relating EM and gravity \cite{Bjerrum-Bohr:2013}.
Furthermore, we have already computed the expansions \eqref{eq:expansion},
hence we only need to compute the leading term in $\Gamma$! Using
the parametrizations \eqref{eq:param}, \eqref{eq:param2}, \eqref{eq:compat}
together with \eqref{eq:k3k4}, we find

\begin{equation}
\begin{split}\Gamma & =(\beta-1)\left(\hat{u}\frac{(1-y)}{2}+v\frac{(1+y)}{2}+(v-\hat{u})\frac{1-y^{2}}{4y}\right)+O(\beta-1)^{2}\,,\end{split}
\label{eq:expansiongamma}
\end{equation}
where
\begin{equation}
\hat{u}=u\left(1-\frac{|\hat{\lambda}][\hat{\lambda}|}{m_{a}}\right)\,.
\end{equation}
We see that the expansion effectively attaches a \textquotedbl{}spin
factor\textquotedbl{} $\left(1-\frac{|\hat{\lambda}][\hat{\lambda}|}{m_{a}}\right)$
to $u$ in the expression \eqref{eq:expansion}. This is expected
since the $A_{(4,i)}^{(S_{a})}$ is built from the 3pt amplitudes
\eqref{eq:3pts}, which can be obtained from the scalar case by promoting
$x_{1}^{h}\rightarrow x_{1}^{h}\left(1-\frac{|\hat{\lambda}][\hat{\lambda}|}{m_{a}}\right)^{S_{a}}$
while $\bar{x}_{1}$ remains the same. Consequently, the expression
\eqref{eq:expansiongamma} precisely reduces to its scalar counterpart
once the spin operator is projected out: Comparing both expansions
we find
\begin{equation}
{\rm Tr}(\Gamma)=2\langle k_{3}|P_{1}|k_{4}]\,,
\end{equation}
as required by \eqref{eq:Gamma}. The conjugation $y\rightarrow-y$
in $\Gamma$ effectively swaps $\tilde{u}\leftrightarrow v$. This
time this transformation also modifies the contribution from the 3pt
amplitudes \eqref{eq:3ptNRLspin}, but once the residue is computed
the leading singularity is still invariant (in the HCL).

Finally, considering the contribution $h_{3}=-h_{4}=-2$ in eq. \eqref{eq:int2}:

\begin{equation}
\begin{split}M_{(S_{a},S_{b},2)}^{(1,b)}=\frac{\beta}{8(\beta^{2}-1)m_{b}^{2}} & \int_{\Gamma_{{\rm LS}}}\frac{dy}{y}A_{4}(P_{1},-P_{2},k_{3}^{-2}(y),k_{4}^{+2}(y))\,\\
 & \otimes A_{3}(P_{3},-L(y),-k_{3}^{+2}(y))A_{3}(-P_{4},L(y),-k_{4}^{-2}(y))\,,
\end{split}
\label{eq:int2-1}
\end{equation}
and inserting \eqref{eq:expansiongamma}, \eqref{eq:expansion}, \eqref{eq:A4spin}
together with \eqref{eq:3ptNRLspin}, we find our second main result
for the classical piece associated to spinning particles

\begin{equation}
\begin{split}M_{(S_{a},S_{b},2)}^{(1,b)} & =\frac{\alpha^{4}}{16}\frac{m_{b}}{\sqrt{-t}(v-u)^{2}}\int_{\infty}\frac{dy}{y^{3}(1-y^{2})^{2}}\left(\hat{u}y(1-y)+vy(1+y)+\left(v-\hat{u}\right)\frac{1-y^{2}}{2}\right)^{\otimes2S_{a}}\,\\
 & \,\quad\times\left(uy(1-y)+vy(1+y)+\frac{(v-u)(1-y^{2})}{2}\right)^{4-2S_{a}}\otimes\left(1-\frac{(1+y)^{2}}{4y}\frac{|\lambda][\lambda|}{m_{b}}\right)^{\otimes S_{b}}
\end{split}
\end{equation}
together with the analogous expression for $|h|=1$. The residue can
then be computed and expanded as a polynomial in spin operators. Evidently,
the factor $\Gamma^{\otimes2S_{a}}$ is responsible for these higher
multipole interactions, together with the spin operators coming from
the 3pt amplitudes \eqref{eq:3ptNRLspin}. Finally, symmetrization
is needed in order to derive the classical potential. This means that
we need to consider the triangle topology obtained by exchanging particles
$m_{a}$ and $m_{b}$. This can be easily done since our expressions
are general as long as $S_{a}$, $S_{b}$ $\leq1$. In appendix \ref{sec:comframe}
we explicitly show how to construct the full answer for $S_{a}=S_{b}=\frac{1}{2}$
in terms of the standard EFT operators, and find full agreement with
the results in \cite{Holstein:08}. This time it can be readily checked
that the Electromagnetic case also leads to analogous spin corrections,
which coincide with those given in \cite{Holstein:2008sw}.

\subsection{Light Bending for Arbitrary Spin}

\label{sub:lightbend}

We will now implement the construction of appendix \ref{sub:Massless-representation}
to obtain the massless limit in a similar fashion as we did for the
scalar case in sec. \ref{sub:Massless-probe-particle}. We will again
focus on the gravitational case since it is of interest for studying
light bending phenomena, addressed in detail in \cite{Bjerrum-Bohr:2016hpa,Bai:2016ivl}
for particles with non trivial helicity.

Let us then proceed to take the massless limit of the parametrization
\eqref{eq:param2} (at $\beta=1$) corresponding to $\tau|\hat{\eta}]\rightarrow0$.
This yields $x_{1}\rightarrow0$, which is in turn equivalent to $u\rightarrow0$.
We get from \eqref{eq:spinexchange}, using \eqref{eq:uv-1}

\begin{equation}
\begin{split}M_{(h_{a},S_{b},2)}^{(0)} & =\alpha^{2}\frac{v^{2}}{t}\left(1-\frac{|\lambda][\lambda|}{m_{b}}\right)^{2S_{b}}\\
 & =\alpha^{2}\dfrac{(s-m_{b}^{2})^{2}}{t}\left(1-\frac{|\lambda][\lambda|}{m_{b}}\right)^{2S_{b}}\,,
\end{split}
\label{eq:spinexchangemassless}
\end{equation}
where $S_{a}=h_{a}$ now corresponds to the helicity of particle $a$.
This operator is to be contracted with the states $|0\rangle$, $|2h_{a}\rangle$
associated to momenta $P_{3}$ and the corresponding ones for $P_{4}$,
which carry the information of the polarizations. It is however trivial
in the sense that it is proportional to the identity for such states,
in particular being independent of $h_{a}$. In the non-relativistic
limit we find $s-m_{b}^{2}\rightarrow2m_{b}E$, with $E\ll m_{b}$
corresponding to the energy of the massless particle. This shows how
the low energy effective potential obtained from \eqref{eq:spinexchangemassless}
is independent of the type of massless particle, as long as it is
minimally coupled to gravity. This is the universality of the light
bending phenomena previously proposed in \cite{Bjerrum-Bohr:2016hpa}.
It may seem that this claim depends on the choice $u=0$ or $v=0$
for defining the massless limit, since for $v=0$ the operator $\left(1-\frac{|\hat{\lambda}][\hat{\lambda}|}{[\hat{\lambda}\hat{\eta}]}\right)^{2h_{a}}$
would certainly show up in the result. However, as argued in the appendix
\ref{sub:Massless-representation}, the choice $v=0$ is supplemented
by the choice of a different basis of states for the massless representation,
such that this operator is again proportional to the identity and
hence independent of $h_{a}$.

To argue for the universality at the 1-loop level, we consider the
massless limit of \eqref{eq:expansiongamma}, given by 
\begin{equation}
\Gamma\rightarrow(\beta-1)\left(v(1+y)+v\frac{1-y^{2}}{2y}\right)\,,
\end{equation}
which is precisely the massless limit of $\langle k_{3}|P_{1}|k_{4}]$,
i.e. the corresponding factor for the scalar case. The conclusion
is that the behavior of $A_{(4,2)}^{(S_{a})}$ is again independent
of $S_{a}=h_{a}$, hence showing the universality. The LS for gravity
now reads

\begin{equation}
M_{(h_{a},S_{b},2)}^{(1,b)}=\left(\frac{\alpha^{4}}{2^{8}}\right)\frac{(s-m_{b})^{2}m_{b}}{\sqrt{-t}}\int_{\infty}\frac{dy\,(1+y)^{6}}{y^{3}(1-y)^{2}}\left(1-\frac{(1+y)^{2}}{4y}\frac{|\lambda][\lambda|}{m_{b}}\right)^{2S_{b}}\,.
\end{equation}
This leading singularity is all what is needed to compute the classical
potential for the massless case, since as explained in subsection
\ref{sub:Massless-probe-particle} the $a-$topology has vanishing
LS. Thus, we note that because there is no need to symmetrize there
is no restriction on $S_{a}$ at all. This means that, up to 1-loop,
we have access to all orders of spin corrections for a massless particle
interacting with a rotating point-like source. The expression can
be used in principle to recover the multipole expansion of the Kerr
black hole solution up to order $G^{2}$, see discussion.

\section{Discussion}

\label{sec:discussion}

In this work we have proposed the implementation of a new set of techniques
in order to extract in a direct manner the classical behavior of a
variety of scattering amplitudes, including arbitrarily high order
spin effects. This classical piece can then be used to construct an
effective field theory for long range gravitational or electromagnetic
interactions. It was shown in \cite{LS} that for the gravitational
case the 1-loop correction to such interaction is completely encoded
into the triangle leading singularity. In this work we have reproduced
this result and extended the argument to the electromagnetic case
in a trivial fashion. The reason this is possible is because the triangle
LS captures the precise non-analytic dependence of the form $t^{-\frac{1}{2}}$,
which carries the subleading contribution to the potential. As explained
in \cite{Holstein:2004dn}, this structure arises from the interplay
between massive and massless propagators entering the loop diagrams.
This is the case whenever massive particles exchange multiple massless
particles which mediate long range forces, such as photons or gravitons.

We have also included the tree level residues for both cases, which
correspond to the leading Newtonian and Coulombian potentials. In
this case, both computations were completely analogous and the gravitational
contribution could be derived by ``squaring'' the electromagnetic
one. This is reminiscent of the double copy construction, which has
been shown to be realized even for the case where massive particles
are involved \cite{Bjerrum-Bohr:2013,Bjerrum-Bohr:2014lea}. At 1-loop
level, such construction is most explicitly realized in the factorization
properties of the Compton amplitude. In the overall picture, this
set of relations between gravity and EM amplitudes renders the computations
completely equivalent. Even though the latter carries phenomenological
interest by itself, it can also be regarded as a model for understanding
long range effects arising in higher PN corrections, including higher
loop and spin orders. 

The HCL was designed as a suitable limit to extract the relevant orders
in $t$ from the complete classical leading singularities introduced
in \cite{LS}. When embedded in this framework, the computation of
the triangle LS proves not only simpler but also leads directly to
$t^{-\frac{1}{2}}$ contribution including all the spin interactions.
As explained in section \ref{sub:HNRL} and explicitly shown in appendix
\ref{sec:comframe}, the covariant form of these interactions allows
us to discriminate them from the purely quantum higher powers of $t$,
which appear merged in the COM frame. In order to distinguish them
we resorted to the following criteria: For a given power of $G$,
a subleading order in $|\vec{q}|$ can be classical if it appears
multiplied by the appropriate power of the spin vector $|\vec{S}|$.
In the HCL framework this is easily implemented since the combination
$|\vec{q}||\vec{S}|$ will always emerge from a covariant form which
does not vanish for $t\rightarrow0$. For instance, for $S=\frac{1}{2}$,
the spin-orbit interaction only arises from $\epsilon_{\alpha\beta\gamma\delta}P_{1}^{\alpha}P_{3}^{\beta}K^{\gamma}S^{\delta}$
and can be tracked directly at leading order.

In striking contrast with previous approaches, the evaluation of spin
effects does not involve increased difficulty with respect to the
scalar case and can be put on equal footing. This is a direct consequence
of implementing the massive representation with spinor helicity variables,
which certainly bypasses all the technical difficulties associated
to the manipulation of polarization tensors. As an important outcome
we have proved that the forms of the higher multipole interactions
are independent of the spin we assign to the scattered particles.
This is a consequence of the equivalence principle, which we have
implemented by assuming minimally coupled amplitudes. The expressions
have been explicitly shown to agree with the previous results in the
literature for the lowest spin orders, corresponding to $S=1$ and
$S=\frac{1}{2}$, yielding spin-orbit, quadrupole and spin-spin interactions.
We emphasize, however, that the proposed expressions correspond to
a relativistic completion of these results, in the sense that they
contain the full $\vec{p}^{\,2}$ expansion.

At this point one could argue that the former difficulty of the diagrammatic
computations has been transferred here to the difficulty in performing
the matching to the EFT operators. In fact, in order to obtain the
effective potential (in terms of vector fields) it is certainly necessary
to translate the spinor helicity operators to their standard forms,
as was done in appendix \ref{sec:spinrep} for $S=\frac{1}{2}$ and
$S=1$. We do not think that this should be regarded as a complication.
First, as a consequence of the universality we have found, it is clear
that we only need to perform the translation once and for a particle
of a given spin $S$, as high as the order of multipole corrections
we require. Second and more important, we think that this work along
with e.g. \cite{Neill:2013wsa,Bjerrum-Bohr:2013,Bjerrum-Bohr:2014zsa,Holstein:2016fxh,AmpforAstro,LS}
will serve as a further motivation towards a complete reformulation
of the EFT framework which naturally integrates recent developments
in scattering amplitudes. For instance, one could aim for a reformulation
of the effective potential, or even better, its replacement by a gauge
invariant observable, solely in terms of spinor helicity variables
so that no translation is needed to address the dynamics of astrophysical
objects.

Next we give some proposals for future work along these lines.

The most pressing future direction is the extension of the leading
singularity techniques in the context of classical corrections at
higher loops \cite{LS}. This is supported by the fact that higher
orders in the PN expansion are associated to characteristic non-analytic
structures in the t channel \cite{Neill:2013wsa}, which are precisely
what the LS captures. By consistency with the PN expansion such higher
orders would require to include spin multipole corrections, so that
both the HCL and the new spin representation emerge as promising additional
tools for such construction. One could hope that with these methods
the scalar and the spinning case will be again on equal footing. Additionally,
the PN expansion also requires to incorporate radiative corrections
and finite-size effects. The latter may be included within the spin
representation presented here by introducing non-minimal couplings,
see e.g. \cite{Levi:2015msa}.

\begin{figure}
\centering\includegraphics[scale=0.5]{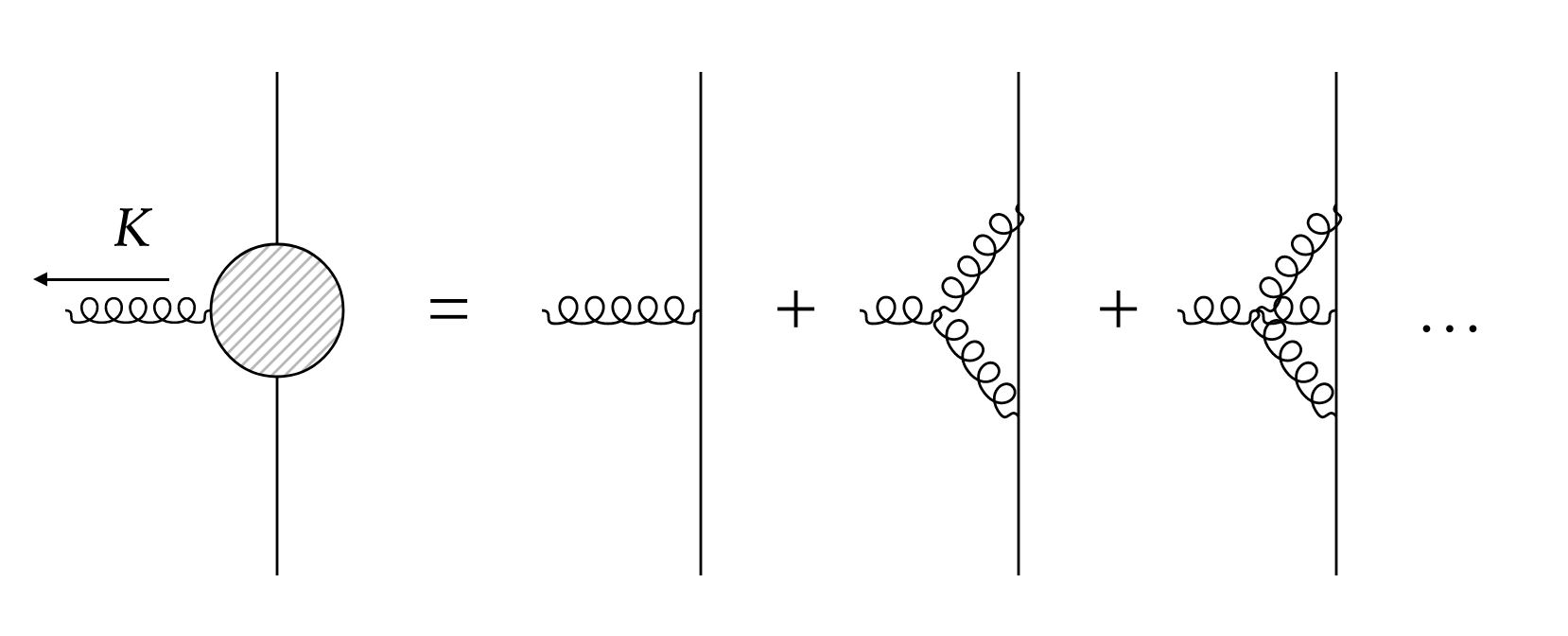}

\caption{The matrix element of the stress-energy tensor $\langle T_{\mu\nu}(K)\rangle$
corresponds to the 3 point function associated to a pair of massive
particles and an external off-shell graviton. The coupling to internal
gravitons emanating from the massive source yields, in the long-range
behavior, higher corrections in $G$. \label{fig:tuv}}
\end{figure}

The first consistency check for higher loop classical corrections
is to reproduce known solutions to Einstein equations. In the spirit
of \cite{Duff,Neill:2013wsa} and the more modern implementations
\cite{Luna:2016hge,Goldberger:2017frp} we could argue that this work
indeed represent progress towards the derivation of classical spacetimes
from scattering amplitudes. As argued by Donoghue \cite{Donoghue:1994dn,Donoghue:2017pgk}
a way to obtain the spacetime metric is to compute the long-range
behavior of the off-shell expectation value $\langle T_{\mu\nu}(K)\rangle$
illustrated in Fig. \ref{fig:tuv}, which yields the Schwarzschild/Kerr
solutions through Einstein equations. At first glance it would seem
that is not possible to compute this matrix element using the on-shell
methods here exposed. However, this is simply analogous to the fact
that we require an off-shell two-body potential for the PN problem.
The solution is, of course, to attach another external particle to
turn Fig. \ref{fig:tuv} into the scattering process of Fig. \ref{fig:intro}.
In this way we can get information about off-shell subprocesses by
examining the $2\rightarrow2$ amplitude. 

A simple way to proceed in that direction is to incorporate probe
particles whose backreaction can be neglected. In fact, the massless
case explored in subsections \ref{sub:Massless-probe-particle} and
\ref{sub:lightbend} can be regarded as a probe particle choice. The
lack of backreaction is realized in the fact that only one triangle
topology is needed for obtaining the classical piece of the amplitude,
which in turn can be thought of containing the process of Fig. \ref{fig:tuv}.
Furthermore, this piece has no restriction in the spin $S$ of the
massive particle, i.e. we can compute both the tree level and 1-loop
potential to arbitrarily high multipole terms. By extracting the matrix
element $\langle T_{\mu\nu}(K)\rangle$ we could recover both leading
and subleading orders in $G$ to arbitrary order in angular momentum
of the Kerr solution, see also \cite{Vaidya:14}. In fact, it was
recently proposed \cite{Vines:2016qwa} that by examining a probe
particle in the Kerr background the generic form of the multipole
terms entering the 2-body Hamiltonian can be extracted at leading
order in $G$ and arbitrary order in spin.

Of course, it is also tempting to explore the opposite direction,
outside the probe particle limit. One could try to obtain an expression
for the effective (i.e. long-range) vertex of Fig. \ref{fig:tuv},
including higher couplings with spin, expressed in terms of spinor
variables. Then an effective potential could be constructed in terms
of several copies of these vertices, for instance to address the n-body
problem in GR \cite{Chu:2008xm,Galaviz:2010te,Hartung:2010jg,Cannella:2011wv}.

\acknowledgments

I would like to thank Freddy Cachazo for proposing this problem and
supervising this work as part of the Perimeter Scholars International
program. I also thank Sebastian Mizera for useful discussions and
comments on the manuscript. I acknowledge the PSI program for providing
me with an academically enriching year, as well as CONICYT for financial
support. Research at Perimeter Institute is supported by the Government
of Canada through Industry Canada and by the Province of Ontario through
the Ministry of Research \& Innovation.

\appendix

\section{Spinor Helicity Variables for Massive Kinematics}

\label{sec:spinrep}

Here we construct the $SU(2)$ states \eqref{eq:reps} and their respective
operators written in terms of anti-chiral spinors, first proposed
in \cite{Nima:16} as a presentation of the massive little group.
In \eqref{eq:reps} we considered two massive particles (with same
mass $m_{b}$ and spin $S$) and constructed the spaces $V_{3}^{S}$,
$\bar{V}_{4}^{S}$ associated to their respective states. We also
introduced the contraction between these states which will naturally
occur in the matrix element of the scattering processes:

\[
\langle n|m\rangle=(-1)^{m}\delta_{m+n,2S}\,.
\]
This follows from the normalization explained in \eqref{eq:normalization}.
It is also natural to define an inner product for each space if we
identify $\bar{V}_{4}^{S}=\left(V_{3}^{S}\right)^{*}$, i.e. as providing
a dual basis for $V_{3}^{S}$ \footnote{The contraction $\langle n|m\rangle$, as defined, is antisymmetric
for fermions. This is reminiscent of the spin-statistics theorem,
as such form is proportional to the minimally coupled 3pt amplitude.
On the other hand, in order to interpret this contraction as an inner
product it is necessary to introduce the dual map $\zeta:V^{S}\rightarrow\left(V^{S}\right)^{*}$.
For instance, defining $\zeta:|n\rangle\mapsto(-1)^{2s}\langle n|$
leads to a symmetric expression.}. With these conventions, we can expand any operator $O\in\left(V_{3}^{S}\right)^{*}\otimes\left(\bar{V}_{4}^{S}\right)^{*}$
as

\begin{equation}
O=\sum_{n,m\leq2S+1}(-1)^{n+m-2S}|\bar{n}\rangle\langle\bar{m}|\,\langle n|O|m\rangle\,,\label{eq:changebasis}
\end{equation}
where $\bar{m}=2S-m,\bar{n}=2S-n$. Of course, this expansion is general
for any choice of basis as long as $|\bar{n}\rangle,\langle\bar{m}|$
are defined as the duals. It is even possible to use different states
for $V_{3}^{S}$ and $V_{4}^{S}$, spanned by different spinors $\{|\lambda],|\eta]\}$
and $\{|\bar{\lambda}],|\bar{\eta}]\}$. However, it is natural to
use the basis \eqref{eq:reps} as it coincides for both massive particles
entering the 3pt amplitude, and also coincides with the dual basis
up to relabelling. Next we explicitly illustrate the natural map between
the states \eqref{eq:reps} and the well known Dirac spinor representation
for $S=\frac{1}{2}$. We also show how to construct the chiral presentation
in terms of angle spinors, in which the basis for both particles turn
out to be different.

First, consider the parametrization \eqref{eq:param}. The basis of
solutions for the (momentum space) Dirac equation are given in terms
of the spinors

\begin{equation}
\begin{split}u_{3}^{+}=\begin{pmatrix}\langle\lambda|\\{}
[\lambda|
\end{pmatrix}\,, & \qquad u_{3}^{-}=\begin{pmatrix}-\langle\eta|\\{}
[\eta|
\end{pmatrix}\,,\\
\\
\bar{u}_{4}^{+}=(-\beta|\lambda\rangle\,\,\ |\lambda])\,, & \qquad\bar{u}_{4}^{-}=(\frac{|\eta\rangle}{\beta}+|\lambda\rangle\,\,\ |\eta])\,.
\end{split}
\label{eq:diracsols}
\end{equation}

(For $\beta=1$, note that \eqref{eq:3ptparam} follows from the Dirac
equation with $x=1$). Thus it is now natural to use $|\eta]$ and
$|\lambda]$ to construct the $S=\frac{1}{2}$ representation for
the (outgoing) particle $P_{4}$, and similarly for $P_{3}$. This
yields an anti-chiral representation of $SU(2)$. From the definition
\eqref{eq:reps} we find (slightly abusing the notation)

\begin{equation}
|+\rangle=\frac{|\lambda]}{\sqrt{m_{b}}}\,,\quad|-\rangle=\frac{|\eta]}{\sqrt{m_{b}}}\,\,\,\in V_{3}^{\frac{1}{2}}\,.\label{eq:normex}
\end{equation}
and analogously for $\langle\pm|\in\bar{V}_{4}^{\frac{1}{2}}$ . The
expansion \eqref{eq:changebasis} leads to the $2\times2$ operator

\begin{equation}
O=\frac{1}{m_{b}}\left(-|\lambda][\lambda|\,O_{(--)}+|\lambda][\eta|\,O_{(-+)}+|\eta][\lambda|\,O_{(+-)}-|\eta][\eta|\,O_{(++)}\right)\,.\label{eq:changebasischiral}
\end{equation}
Had we used the chiral part, we would have selected a different basis
for each of the massive particles. In fact, the chiral part is obtained
by acting with $P_{3}$, $P_{4}$ on the anti-chiral states, respectively.
This means that the change of basis (for $S=\frac{1}{2}$) is given
by

\begin{equation}
\bar{O}=\frac{\bar{P}_{3}OP_{4}}{m^{2}}\,,\label{eq:changebasisantichiral}
\end{equation}
where we have used matrix multiplication, with the extension to higher
values of spin being straightforward. 

For completeness we present here some useful expressions obtained
at $\beta=1$, even though they can easily be computed in general

\begin{equation}
\begin{split}m^{2}\bar{u}_{4}\gamma_{\mu}u_{3}\rightarrow m^{2}\gamma_{\mu} & =2(P_{4})_{\mu}|\eta][\lambda|-2(P_{3})_{\mu}|\lambda][\eta|-2v_{\mu}|\lambda][\lambda|\\
 & =2m(P_{3})_{\mu}+2K_{\mu}|\eta][\lambda|-2v_{\mu}|\lambda][\lambda|\,,\\
\bar{u}_{4}u_{3}\rightarrow\mathbb{I}_{2\times2} & =\dfrac{(P_{3})^{\mu}}{m}\gamma_{\mu}=2-\frac{|\lambda][\lambda|}{m}\,,\\
\frac{m^{2}}{2}\bar{u}_{4}\gamma_{5}\gamma_{\mu}u_{3}\rightarrow m^{2}S_{\mu} & =2K_{\mu}|\eta][\eta|-2(R_{\mu}+\frac{1}{2}v_{\mu})|\lambda][\lambda|\\
 & \qquad+2(u_{\mu}-v_{\mu}+K_{\mu})|\eta][\lambda|+2(u_{\mu}-v_{\mu})|\lambda][\eta|\,,
\end{split}
\label{eq:translation}
\end{equation}
where 
\begin{equation}
\begin{split}2v_{\mu}=[\eta|\sigma_{\mu}|\lambda\rangle\,, & \quad2u_{\mu}=[\lambda|\sigma_{\mu}|\eta\rangle\,,\\
v_{\mu}+u_{\mu}=(P_{3})_{\mu}\,, & \quad2R_{\mu}=[\eta|\sigma_{\mu}|\eta\rangle\,.
\end{split}
\label{eq:vecuv}
\end{equation}
Here $\mathbb{I}_{2\times2}$ is the identity operator for Dirac spinors,
projected into the two-dimensional subspaces spanned by the wavefunctions
$u^{\pm}$. On the other hand, in the second line we used the identity
\begin{equation}
1=\dfrac{|\eta][\lambda|-|\lambda][\eta|}{[\lambda\eta]}\,.\label{eq:spinid}
\end{equation}

From the fourth line of \eqref{eq:translation}, using $2q\cdot K=-m^{2}$
we find in the HCL

\begin{equation}
S_{\mu}K^{\mu}=|\lambda][\lambda|\,.\label{eq:spinop-1}
\end{equation}
This is the reason we call $|\lambda][\lambda|$ a spin operator.
One may wonder why the spin operator appears in the expansion of $\mathbb{I}_{2\times2}$,
which contains the scalar contribution. Even though $\mathbb{I}$
and $\gamma_{5}\gamma_{\mu}$ are orthogonal as Dirac matrices, this
does not hold once they are projected into the 2D subspace of physical
states. This is consistent with the non-relativistic expansions of
\cite{Holstein:08}, where the form $\bar{u}_{4}u_{3}$ indeed contributes
to the spin interaction. In fact, this is also true for higher spin
generalizations as we now show.

Motivated by the manifest universality found in section \ref{sec:spin},
i.e. expression \eqref{eq:spinexchange}, we consider the following
extensions for arbitrary spin $S_{b}$ (not to be confused with the
spin vector $S_{\mu}$)
\begin{eqnarray}
S_{\mu}K^{\mu} & = & 2S_{b}|\lambda][\lambda|\,,\label{eq:claim}\\
\mathbb{I}_{(2S_{b}+1)} & = & 2\left(1-S_{b}\frac{|\lambda][\lambda|}{m}\right)\,,\nonumber 
\end{eqnarray}
As explained in the discussion after Eq. \eqref{eq:3pts}, we omit
the trivial part of the operators on the RHS. This allows to keep
the expressions compact and makes the universality manifest. Let us
briefly perform a nontrivial check of equations \eqref{eq:claim}
for higher spins. We do so by examining the representation for $S_{b}=1$,
which in the standard framework is given by polarization vectors satisfying
$\epsilon^{(i)}\cdot P=0$, $i=1,2,3$, for a given momentum $P^{2}=m_{b}^{2}$.
In terms of spinor helicity variables the polarization vectors $\epsilon_{3}$
and $\epsilon_{4}$ are represented as operators acting on $V_{3}^{1}$
and $\bar{V}_{4}^{1}$ respectively. Explicitly, we can choose \footnote{Here we use the notation $[\lambda|[\eta|$ to account for the standard
tensor product, i.e. not symmetrized. Of course, we can replace $[\lambda|[\eta|\rightarrow\frac{1}{\sqrt{2}}[\lambda|\odot[\eta|$,
where $\odot$ involves the normalization \eqref{eq:normalization}. }

\begin{eqnarray*}
\frac{m_{b}^{2}(\epsilon_{3})_{\mu}}{2} & \rightarrow & [\lambda|[\lambda|R_{\mu}-[\lambda|[\eta|(u-v+K)_{\mu}-[\eta|[\eta|K_{\mu}\,,\\
\frac{m_{b}^{2}(\epsilon_{4})_{\mu}}{2} & \rightarrow & |\lambda]|\lambda](R+\frac{1}{2}P_{3})_{\mu}-|\lambda]|\eta](u-v-K)_{\mu}-|\eta]|\eta]K_{\mu}\,,
\end{eqnarray*}
Using this expression it is easy to check the validity of \eqref{eq:claim}
for $S_{b}=1$, with \cite{Holstein:08} 
\begin{eqnarray}
\epsilon_{3}\cdot\epsilon_{4} & \rightarrow & \mathbb{\mathbb{I}}_{3}\,,\nonumber \\
\frac{1}{2m_{b}}\epsilon_{\mu\alpha\beta\gamma}\epsilon_{4}^{\alpha}\epsilon_{3}^{\beta}(P_{3}+P_{4})^{\gamma} & \rightarrow & S_{\mu}\,.\label{eq:spin1op}
\end{eqnarray}
Also, we can now derive the form of the quadrupole interaction. It
is given by 
\begin{equation}
(\epsilon_{4}\cdot K)(\epsilon_{3}\cdot K)=|\lambda]|\lambda]\otimes[\lambda|[\lambda|\,.\label{eq:quadrupole}
\end{equation}
We will use this expression in appendix \ref{sec:comframe} to translate
the leading singularity into standard EFT operators.

For illustration purposes, let us close this section by constructing
the representation of the 3pt amplitudes for $S=\frac{1}{2}$ massive
fields with a graviton. Let the polarization of the massless particle
be described by $|\bar{\lambda}]=\sqrt{x}|\lambda]$, $\langle\bar{\lambda}|=\frac{\langle\lambda|}{\sqrt{x}}$,
where $x$ carries helicity $1$ (recall $|\lambda]$ is fixed) and
agrees with \eqref{eq:3ptparam}. The 3pt amplitude is given by \cite{Vaidya:14}

\begin{equation}
\begin{split}A_{\frac{1}{2}}^{(+2)} & =\frac{\alpha m}{2}\gamma_{\mu}\dfrac{[\bar{\lambda}|\sigma^{\mu}|\eta\rangle[\bar{\lambda}|P_{3}|\eta\rangle}{\langle\eta\bar{\lambda}\rangle^{2}}\,,\\
A_{\frac{1}{2}}^{(-2)} & =\frac{\alpha m}{2}\gamma_{\mu}\dfrac{[\eta|\sigma^{\mu}|\bar{\lambda}\rangle[\eta|P_{3}|\bar{\lambda}\rangle}{[\eta\bar{\lambda}]^{2}}\,.
\end{split}
\end{equation}
Here we have fixed the reference spinor entering in the 3pt. amplitudes
to be $\eta$. Using \eqref{eq:translation} together with \eqref{eq:param}
we find

\begin{equation}
\begin{split}A_{\frac{1}{2}}^{(+2)} & =\alpha(mx)^{2}\left(1-\frac{|\lambda][\lambda|}{m}\right)\,,\\
A_{\frac{1}{2}}^{(-2)} & =\alpha\left(\dfrac{m}{x}\right)^{2}\,,
\end{split}
\label{eq:appt}
\end{equation}
precisely agreeing with \eqref{eq:3pts} for $|h|=2$. Furthermore,
in the chiral representation we find, using \eqref{eq:changebasisantichiral}

\begin{equation}
\begin{split}\bar{A}{}_{\frac{1}{2}}^{(+2)} & =\alpha\left(\frac{m}{\bar{x}}\right)^{2}\,,\\
\bar{A}_{\frac{1}{2}}^{(-2)} & =\alpha(m\bar{x})^{2}\left(1-\frac{|\lambda\rangle\langle\lambda|}{m}\right)\,.
\end{split}
\end{equation}
where $\bar{x}$ is defined in \eqref{eq:3ptparam}.\\

\subsection{Massless Representation\label{sub:Massless-representation}}

We can extend the treatment described in section \ref{sub:Massless-probe-particle}
in order to construct the states for massless particles. The idea
is to use the two highest weight states $|0\rangle$, $|2S\rangle$
of the massive representation as the physical polarizations of the
massless one, after the limit is taken. The massless case can be formally
defined by introducing a variable $\tau$ in the parametrization \eqref{eq:param},
i.e. by putting either $|\eta]\mapsto\tau|\eta]$ or $|\eta\rangle\mapsto\tau|\eta\rangle$
and then proceed to take the limit $\tau\rightarrow0$. This parametrizes
the mass of both $P_{3}(\tau)$ and $P_{4}(\tau)$ as $m^{2}(\tau)=\tau m^{2}$.
Next we proceed to sketch the procedure leading to the massless 3pt.
amplitudes\footnote{At this level we keep the discussion general for $S$ and $h$. Of
course, (interacting) massless higher spin particles are known to
be inconsistent by very fundamental principles, thus effectively restricting
our choices to $S,h\leq2$. } and study both choices $\tau|\eta]\rightarrow0$ and $\tau|\eta\rangle\rightarrow0$.
As these amplitudes correspond to the building blocks for both the
tree level residue and the triangle LS in section \ref{sec:spin},
showing that they can be recovered from our expressions \eqref{eq:3pts}
proves the equivalence with the standard spinor helicity approach
for massless particles. This approach was recently implemented in
\cite{Bjerrum-Bohr:2016hpa}.

In the following we will consider $\beta=1$. Indeed, the massless
deformation of the momenta is only consistent in the HCL since $t=\tau\frac{(\beta-1)^{2}}{\beta}m_{b}^{2}$$\rightarrow0$
as $\tau\rightarrow0$. This is enough for our purposes in section
\ref{sec:spin} since we evaluate both the tree level residue and
triangle LS by neglecting subleading contributions in $t$. For the
choice $|\eta]\mapsto\tau|\eta]$ we thus have
\begin{equation}
\begin{split}P_{3} & =\tau|\eta]\langle\lambda|+|\lambda]\langle\eta|\,\longrightarrow|\lambda]\langle\eta|\,,\\
P_{4} & =\tau|\eta]\langle\lambda|+|\lambda](\langle\eta|+\langle\lambda|)\,\longrightarrow|\lambda](\langle\eta|+\langle\lambda|)\,,\\
K & =|\lambda]\langle\lambda|\,.
\end{split}
\label{eq:massless1}
\end{equation}
In the following we choose $|\lambda]$,$\langle\lambda|$ to represent
the polarizations of the particle $K$. As explained in section \eqref{sub:Massless-probe-particle},
it is convenient to reabsorb the mass into the definition of $x$
\eqref{eq:3ptparam}, thus we have

\begin{equation}
x=\tau[\lambda\eta]=\tau m\,,\quad\bar{x}=\langle\lambda\eta\rangle=m\,.
\end{equation}
This means $\tau|\eta]\rightarrow0$ is equivalent to the limit $x\rightarrow0$,
keeping $\bar{x}$ fixed. As the reference spinor $|\eta]$ is also
fixed, we can assume that the neither the basis \eqref{eq:reps} nor
the operators \eqref{eq:3pts} depend on $\tau$ in any other way
that is not through $x$. With these considerations, we find for the
massless limit 
\begin{equation}
\begin{split}A_{S}^{(h)}=0\,, & \quad A_{S}^{(-h)}=\alpha\bar{x}^{h}\,,\end{split}
\label{eq:3ptslim}
\end{equation}
where at this stage $\bar{x}=\langle\lambda\eta\rangle$ is not restricted
since the original mass $m$ is not relevant after the limit is taken.
We then note that all the positive helicity amplitudes vanish. In
fact, these amplitudes can be described in terms of square brackets,
thus it is expected that they vanish for the $\tau=0$ limit in \eqref{eq:massless1}.
Now, the negative helicity amplitudes in the standard spinor helicity
notation read \cite{Elvang}

\begin{eqnarray}
A(3^{+S},4^{-S},K^{-h}) & = & \alpha\frac{\langle K3\rangle^{h-2S}\langle K4\rangle^{h+2S}}{\langle43\rangle^{h}}\label{eq:3ptmasslesss}\\
 & = & \alpha\bar{x}^{h}\,.\nonumber 
\end{eqnarray}
Note that this derivation is also valid for $A(3^{-S},4^{+S},K^{-h})$
up to a possible sign. Also, the configuration $A(3^{+S},4^{+S},K^{-h})$
together with its conjugate do not correspond to the minimal coupling
and thus vanish. In order to interpret these amplitudes as matrix
elements of \eqref{eq:3ptslim}, we need to specify the basis of states
for the massless particles. It turns out that just the highest weight
states in \eqref{eq:reps} are enough for this purpose. That is, we
find
\begin{eqnarray}
A(3^{+S},4^{-S},K^{-h})=\langle2S|A_{S}|0\rangle & , & A(3^{-S},4^{+S},K^{-h})=\langle0|A_{S}|2S\rangle\,,\label{eq:expval}\\
A(3^{+S},4^{+S},K^{-h})=\langle2S|A_{S}|2S\rangle & , & A(3^{+S},4^{+S},K^{-h})=\langle2S|A_{S}|2S\rangle\,,\nonumber 
\end{eqnarray}
therefore showing the equivalence of both approaches for massless
particles. Here we note that a somehow more straightforward approach
is to define the massless limit directly in the expectation values
\eqref{eq:expval}, following \cite{Nima:16}. Instead, we have opted
for constructing the corresponding operators \eqref{eq:3ptslim},
since our integral expressions in section \ref{sec:spin} are given
in terms of them. These operators are defined for the basis built
from the fixed spinors $|\lambda]$ and $|\eta]$, which are reminiscent
of the massive representation in \eqref{eq:massless1}.

The choice $|\eta\rangle\mapsto\tau|\eta\rangle$ is completely analogous
and yields

\begin{equation}
\begin{split}A_{S}^{(h)}=\alpha x^{h}\left(1-\frac{|\lambda][\lambda|}{[\lambda\eta]}\right)^{S}\,, & \quad A_{S}^{(-h)}=0\end{split}
\,,\label{eq:3ptslim-1}
\end{equation}
i.e. vanishing negative helicity amplitudes. This is expected since
we have

\begin{equation}
\begin{split}P_{3} & =|\eta]\langle\lambda|+\tau|\lambda]\langle\eta|\,\longrightarrow|\eta]\langle\lambda|\,,\\
P_{4} & =|\eta]\langle\lambda|+\tau|\lambda]\langle\eta|+|\lambda]\langle\lambda|\,\longrightarrow(|\lambda]+|\eta])\langle\lambda|\,.
\end{split}
\label{eq:massless1-1}
\end{equation}
However, this time we note that the natural basis of spinors for $P_{4}$
is given by $|\bar{\eta}]:=|\lambda]+|\eta]$ and $|\lambda]$. When
expressed in terms of this basis, the expression \eqref{eq:3ptslim-1}
takes a form analogous to \eqref{eq:3ptslim}. Hence we construct
the states $\langle0|$,$\langle2S|$ in $\bar{V}_{4}^{S}$ in terms
of these spinors.

\section{Matching the Spin Operators}

\label{sec:comframe}

Here we explain how to recover the standard form of the potential
in terms of generic spin operators \eqref{eq:covmultipole}, starting
from the results of section \ref{sec:spin}. As usual throughout this
work, we focus on the gravitational case since it presents greater
difficulty in the standard approaches. We give two examples which
illustrate how the procedure works. First, we present the tree level
result for the case $S_{a}=0$, $S_{b}=1$, which yields both a spin-orbit
and a quadrupole interaction. Second, we discuss the matching at 1-loop
level for the case $S_{a}=S_{b}=\text{\ensuremath{\frac{1}{2}}}$.
Both computations were done in \cite{Holstein:08} using standard
Feynman diagrammatic techniques, which lead to notably increased difficulty
with respect to the scalar case. Here we find that the computations
are straightforward using the techniques introduced throughout this
work. In fact, all the computations in \cite{Holstein:08} can be
redone in a direct way and we leave them as an exercise for the reader.
The same can be done for the EM case in order to recover the results
previously presented in \cite{Holstein:2008sw}.

The starting point for both cases are the explicit expressions for
the variables $u,v$ that we used to construct the amplitudes. We
can easily solve them from Eq. \eqref{eq:uv-1}. We find 
\begin{eqnarray}
2u & = & s-m_{a}^{2}-m_{b}^{2}+\sqrt{\left(s-m_{a}^{2}-m_{b}^{2}\right)^{2}-4m_{a}^{2}m_{b}^{2}}\,,\label{eq:uvexplicit}\\
2v & = & s-m_{a}^{2}-m_{b}^{2}-\sqrt{\left(s-m_{a}^{2}-m_{b}^{2}\right)^{2}-4m_{a}^{2}m_{b}^{2}}\,,\nonumber 
\end{eqnarray}
where the square root corresponds to the parity odd piece. From the
definition \eqref{eq:uvdef} it is clear that under the exchange $P_{1}\leftrightarrow P_{3}$
(which we perform below), $u$ and $v$ must also be exchanged. Now,
to keep the notation compact, let us write
\[
P_{1}\cdot P_{3}=rm_{a}m_{b}\,,\quad r>1.
\]
Note that in the non-relativistic regime we have $r\rightarrow1$.
Now we can write \ref{eq:uvexplicit} as
\begin{eqnarray}
u & =m_{a}m_{b}\left(r+\sqrt{r^{2}-1}\right)\,,\quad & v=m_{a}m_{b}\left(r-\sqrt{r^{2}-1}\right)\,.\label{eq:uvcompact}
\end{eqnarray}
 Consider now the case $S_{a}=0$, $S_{b}=1$. Let us construct a
linear combination of the EFT operators associated to scalar, spin-orbit,
and quadrupole interaction, that is \cite{Holstein:08,Vaidya:14}

\begin{equation}
\bar{M}_{(0,1,2)}^{(1)}=\alpha^{2}\dfrac{(m_{a}m_{b})^{2}}{t}\left(c_{1}(r)\epsilon_{3}\cdot\epsilon_{4}+c_{2}(r)\frac{\epsilon_{\mu\alpha\beta\gamma}K^{\mu}P_{1}^{\alpha}P_{3}^{\beta}S^{\gamma}}{m_{a}m_{b}^{2}}+c_{3}(r)\frac{(\epsilon_{4}\cdot K)(\epsilon_{3}\cdot K)}{m_{b}^{2}}\right)\,.\label{eq:ansatz}
\end{equation}
The reason we call $\epsilon_{3}\cdot\epsilon_{4}$ a scalar interaction
is because, as will be evident in a moment, it is the only piece surviving
the contraction $\langle0|\bar{M}_{(0,1,2)}^{(1)}|2\rangle$, which
we identified as the scalar amplitude (see discussion below Eq. \eqref{eq:spinexp}). 

Note that we have not assumed the non-relativistic limit in the $u,v$
variables, only the HCL $t=0$ which selects the classical contribution.
The operators can now be expanded using \eqref{eq:claim}, \eqref{eq:quadrupole}.
For this, it is enough to note that in the HCL the spin-orbit piece
takes the form

\begin{equation}
\epsilon_{\mu\alpha\beta\gamma}K^{\mu}P_{1}^{\alpha}P_{3}^{\beta}S^{\gamma}=\frac{K\cdot S}{2}\sqrt{\left(s-m_{a}^{2}-m_{b}^{2}\right)^{2}-4m_{a}^{2}m_{b}^{2}}=m_{a}m_{b}(K\cdot S)\sqrt{r^{2}-1}\,.\label{eq:dethcl}
\end{equation}
We then find

\[
\bar{M}_{(0,1,2)}^{(0)}=\alpha^{2}\dfrac{(m_{a}m_{b})^{2}}{t}\left(2c_{1}-2\frac{|\lambda][\lambda|}{m_{b}}(c_{1}-c_{2}\sqrt{r^{2}-1})+c_{3}\frac{|\lambda]|\lambda]\otimes[\lambda|[\lambda|}{m_{b}^{2}}\right)\,.
\]
Comparing now with the expression \eqref{eq:spinexchange}, which
in this case reads 

\begin{eqnarray*}
M_{(0,1,2)}^{(0)} & = & \dfrac{\alpha^{2}}{t}\left(u^{2}+v^{2}\left(1-\frac{|\lambda][\lambda|}{m_{b}}\right)^{2}\right)\\
 & = & \dfrac{\alpha^{2}}{t}\left(u^{2}+v^{2}-2v^{2}\frac{|\lambda][\lambda|}{m_{b}}+v^{2}\frac{|\lambda]|\lambda]\otimes[\lambda|[\lambda|}{m_{b}^{2}}\right)\\
 & = & \alpha^{2}\dfrac{(m_{a}m_{b})^{2}}{t}\left((4r^{2}-2)-2\left(2r^{2}-1-2r\sqrt{r^{2}-1}\right)\frac{|\lambda][\lambda|}{m_{b}}\right.\\
 &  & \left.+\left(2r^{2}-1-2r\sqrt{r^{2}-1}\right)\frac{|\lambda]|\lambda]\otimes[\lambda|[\lambda|}{m_{b}^{2}}\right)\,,
\end{eqnarray*}

we find 
\begin{eqnarray*}
c_{1} & = & 2r^{2}-1\,,\\
c_{2} & = & 2r\,,\\
c_{3} & = & 2r^{2}-1-2r\sqrt{r^{2}-1}\,.
\end{eqnarray*}
The result in \cite{Holstein:08} can then be recovered by imposing
the non-relativistic limit $s\rightarrow s_{0}$, which in this case
reads $r\rightarrow1$\footnote{There are, however, some discrepancies in conventions which may be
fixed by replacing $-\epsilon_{f}^{b*}\rightarrow\epsilon_{4}$, $iS_{b}\rightarrow S_{b}$
in \cite{Holstein:08}. We find our conventions more appropriated
since the sign in the scalar interaction is the same for any spin.}. Even though we computed the residue in \eqref{eq:ansatz} at $t=0$,
it is evident that this expression can be analytically extended to
the region $t\neq0$ in which the COM frame can be imposed, as described
in \eqref{sub:HNRL}. This is precisely done in \cite{Holstein:08}
where the effective potential is obtained from this expression after
implementing the Born approximation.

The 1-loop result for $S_{a}=0$, $S_{b}=1$ can be computed in the
same fashion, by using the expressions provided in section \eqref{sub:spin1loop}.
Expectedly, the EFT operators are exactly the same that appeared at
tree level, but the behavior of the coefficients $c_{1}$, $c_{2}$
and $c_{3}$ as functions of $r$ differs in the sense that it can
involve poles at $r=1$. We now illustrate all this by considering
the more complex case also addressed in \cite{Holstein:08}, namely
$S_{a}=S_{b}=\frac{1}{2}$.

For $S=\frac{1}{2}$ the multipole operators are restricted to the
scalar and spin-orbit interaction. They read \cite{Holstein:08}
\begin{eqnarray*}
\mathcal{U}=\bar{u}_{4}u_{3} & ,\, & \mathcal{E}=\epsilon_{\alpha\beta\gamma\delta}P_{1}^{\alpha}P_{3}^{\beta}K^{\gamma}S^{\delta}\,.
\end{eqnarray*}
In our case we will consider two copies of these operators, one for
each particle. That is to say we propose the following form for the
1-loop leading singularity

\begin{eqnarray}
\bar{M}_{(\frac{1}{2},\frac{1}{2},2)}^{(1)} & = & \left(\frac{\alpha^{4}}{16}\right)\frac{\left(m_{a}m_{b}\right)^{2}}{\sqrt{-t}}\left(c_{11}\mathcal{U}_{a}\mathcal{U}_{b}+c_{12}\frac{\mathcal{U}_{a}\mathcal{E}_{b}}{m_{b}^{2}m_{a}}+c_{21}\frac{\mathcal{E}_{a}\mathcal{U}_{b}}{m_{a}^{2}m_{b}}+c_{22}\frac{\mathcal{E}_{a}\mathcal{E}_{b}}{m_{b}^{3}m_{a}^{3}}\right)\,\label{eq:ansatzspin1/2}\\
 & = & \alpha^{4}\frac{\left(m_{a}m_{b}\right)^{2}}{4\sqrt{-t}}\left(c_{11}-\frac{c_{11}-c_{12}\sqrt{r^{2}-1}}{2}\left(\frac{|\hat{\lambda}][\hat{\lambda}|}{m_{a}}\right)-\frac{c_{11}+c_{21}\sqrt{r^{2}-1}}{2}\left(\frac{|\lambda][\lambda|}{m_{b}}\right)\right.\nonumber \\
 &  & \left.+\frac{\left(c_{11}-(c_{12}-c_{21})\sqrt{r^{2}-1}-c_{22}(r^{2}-1)\right)}{4}\frac{|\hat{\lambda}][\hat{\lambda}|}{m_{a}}\otimes\frac{|\lambda][\lambda|}{m_{b}}\right)\,.\nonumber 
\end{eqnarray}
Here we have used \eqref{eq:dethcl},\eqref{eq:spinop-1} and \eqref{eq:translation}.
A minus sign was introduced when implementing \eqref{eq:spinop-1}
for particle $m_{a}$, which arises from the mismatch between both
parametrizations in the HCL, i.e. $K=|\lambda]\langle\lambda|=-|\hat{\lambda}][\hat{\lambda}|$.
We proceed to compare this with the sum of the two triangle leading
singularities given by \eqref{eq:int2}, using the results of section
\ref{sub:spin1loop}. The result can be written

\[
M_{(\frac{1}{2},\frac{1}{2},2)}^{(1,{\rm full})}=M_{(\frac{1}{2},\frac{1}{2},2)}^{(1,b)}+M_{(\frac{1}{2},\frac{1}{2},2)}^{(1,a)}\,,
\]
where $M_{(\frac{1}{2},\frac{1}{2},2)}^{(1,a)}$ is obtained by exchanging
$m_{a}\leftrightarrow m_{b}$, $|\hat{\lambda}][\hat{\lambda}|\leftrightarrow|\lambda][\lambda|$
and $u\leftrightarrow v$ in

\begin{equation}
\begin{split}M_{(\frac{1}{2},\frac{1}{2},2)}^{(1,b)} & =\left(\frac{\alpha^{4}}{16}\right)\frac{m_{b}}{\sqrt{-t}(v-u)^{2}}\int_{\infty}\frac{dy}{y^{3}(1-y^{2})^{2}}\left(\hat{u}y(1-y)+vy(1+y)+\left(v-\hat{u}\right)\frac{1-y^{2}}{2}\right)\,\\
 & \,\quad\quad\times\left(uy(1-y)+vy(1+y)+\frac{(v-u)(1-y^{2})}{2}\right)^{3}\otimes\left(1-\frac{(1+y)^{2}}{4y}\frac{|\lambda][\lambda|}{m_{b}}\right)\,,
\end{split}
\label{eq:3ptNRLspin-1}
\end{equation}
with $\hat{u}=u\left(1-\frac{|\hat{\lambda}][\hat{\lambda}|}{m_{a}}\right)$.
After computing the contour integral, we can easily solve for the
coefficients $c_{ij}$, $i,j\in\{1,2\}$. In order to compare with
the results in the literature, we first perform the non-relativistic
expansion

\begin{eqnarray}
c_{11} & = & 6(m_{a}+m_{b})+\ldots\nonumber \\
c_{12} & = & \frac{4m_{a}+3m_{b}}{2(r-1)}+11\left(m_{a}+\frac{3}{4}m_{b}\right)+\ldots\nonumber \\
c_{21} & = & \frac{3m_{a}+4m_{b}}{2(r-1)}+11\left(\frac{3}{4}m_{a}+m_{b}\right)+\ldots\label{eq:coefs}\\
c_{22} & = & \frac{m_{a}+m_{b}}{4(r-1)^{2}}+\frac{9(m_{a}+m_{b})}{2(r-1)}+\ldots\nonumber 
\end{eqnarray}
Note that even though the coefficients present poles, they are parity
invariant in the sense that they do not contain square roots. To put
the result in the same form as \cite{Holstein:08}, we need to further
extract the standard spin-spin interaction term from our operator
$\mathcal{E}_{a}\mathcal{E}_{b}$. This accounts for extracting the
classical piece, which can be obtained by returning to the physical
region $t=K^{2}\neq0$. Using \eqref{eq:spinid} we find
\[
\mathcal{E}_{a}\mathcal{E}_{b}=m_{a}m_{b}(r^{2}-1)\left((S_{a}\cdot K)(S_{b}\cdot K)-K^{2}(S_{a}\cdot S_{b})\right)+rK^{2}(P_{1}\cdot S_{b})(P_{3}\cdot S_{a})+O(K^{3})\,,
\]
where $O(K^{3})=O(|\vec{q}|^{3})$ denotes quantum contributions,
i.e. higher orders in $|\vec{q}|$ for a fixed power of spin $|\vec{S}|$.
However, we note the presence of the couplings $P_{i}\cdot S\sim\vec{v}\cdot\vec{S}$
which certainly do not appear in the effective potential \cite{Barker1979,Holstein:08,Vaidya:14}.
In fact, in the standard EFT framework these couplings are dropped
by the so-called Frenkel-Pirani conditions or Tulczyjew conditions
\cite{Frob:2016xte}\footnote{They can arise, however, when including non-minimal couplings corresponding
to higher dimensional operators, see e.g. \cite{Levi:2015msa}}. In our case they have emerged due to our bad choice of ansatz \eqref{eq:ansatzspin1/2}.
In fact, the right choice is now clearly obtained by replacing 
\[
\mathcal{E}_{a}\mathcal{E}_{b}\rightarrow m_{a}m_{b}(r^{2}-1)\left((S_{a}\cdot K)(S_{b}\cdot K)-K^{2}(S_{a}\cdot S_{b})\right)\,,
\]
corresponding to the correct spin-spin interaction term \cite{Porto:2005ac},
which is already visible at tree level \cite{Holstein:2008sw,Holstein:08,Vaidya:14}.
Note, however, that this does not modify the HCL of this operator,
which comes solely from the first term. Consequently, our results
\eqref{eq:coefs} are still valid and indeed they agree with the ones
in the literature \cite{Holstein:08}. They can be regarded as a fully
relativistic completion leading to the effective potential up to order
$G^{2}$.

\bibliographystyle{JHEP}
\bibliography{references}

\end{document}